\documentclass[12pt,a4paper,english,amssymb,nofootinbib,superscriptaddress]{revtex4}
\usepackage{lmodern}
\usepackage{lmodern}
\usepackage[T1]{fontenc}
\usepackage[latin9]{inputenc}
\usepackage{color}
\usepackage{babel}
\usepackage{verbatim}
\usepackage{amsmath}
\usepackage{amssymb}
\usepackage{esint}
\usepackage[unicode=true,pdfusetitle,
 bookmarks=true,bookmarksnumbered=false,bookmarksopen=false,
 breaklinks=false,pdfborder={0 0 1},backref=false,colorlinks=true]
 {hyperref}
\hypersetup{
 citecolor=blue,urlcolor=blue,linkcolor=blue}

\makeatletter

\pdfpageheight\paperheight
\pdfpagewidth\paperwidth

\@ifundefined{textcolor}{}
{%
 \definecolor{BLACK}{gray}{0}
 \definecolor{WHITE}{gray}{1}
 \definecolor{RED}{rgb}{1,0,0}
 \definecolor{GREEN}{rgb}{0,1,0}
 \definecolor{BLUE}{rgb}{0,0,1}
 \definecolor{CYAN}{cmyk}{1,0,0,0}
 \definecolor{MAGENTA}{cmyk}{0,1,0,0}
 \definecolor{YELLOW}{cmyk}{0,0,1,0}
}

\usepackage{babel}
\@ifundefined{definecolor}{color}{}
\usepackage{babel}
\@ifundefined{definecolor}{color}{}
\@ifundefined{definecolor}{color}{}
\usepackage{babel}
\@ifundefined{definecolor}{color}{}
\usepackage{babel}
\pdfpageheight\paperheight
\pdfpagewidth\paperwidth

\@ifundefined{textcolor}{}{%
 \definecolor{BLACK}{gray}{0}
 \definecolor{WHITE}{gray}{1}
 \definecolor{RED}{rgb}{1,0,0}
 \definecolor{GREEN}{rgb}{0,1,0}
 \definecolor{BLUE}{rgb}{0,0,1}
 \definecolor{CYAN}{cmyk}{1,0,0,0}
 \definecolor{MAGENTA}{cmyk}{0,1,0,0}
 \definecolor{YELLOW}{cmyk}{0,0,1,0}
 }

\@ifundefined{definecolor}{color}{}
\usepackage{latexsym}\usepackage{bm}

\usepackage{amssymb}

\newcommand*{\QEDA}{\hfill\ensuremath{\blacksquare}}%

\makeatother

\begin{document}

\title{Kerr-Schild--Kundt Metrics are Universal}

\author{Metin G{\" u}rses}

\email{gurses@fen.bilkent.edu.tr}

\selectlanguage{english}%

\affiliation{{\small{}Department of Mathematics, Faculty of Sciences}\\
 {\small{}Bilkent University, 06800 Ankara, Turkey}}

\author{Tahsin \c{C}a\u{g}r\i{} \c{S}i\c{s}man}

\email{tahsin.c.sisman@gmail.com}

\selectlanguage{english}%

\affiliation{Department of Astronautical Engineering,\\
 University of Turkish Aeronautical Association, 06790 Ankara, Turkey}

\author{Bayram Tekin}

\email{btekin@metu.edu.tr}

\selectlanguage{english}%

\affiliation{Department of Physics,\\
 Middle East Technical University, 06800 Ankara, Turkey}
\begin{abstract}
We define (non-Einsteinian) universal metrics as the metrics that
solve the source-free covariant field equations of generic gravity
theories. Here, extending the rather scarce family of universal metrics
known in the literature, we show that the Kerr-Schild--Kundt class
of metrics are universal. Besides being interesting on their own,
these metrics can provide consistent backgrounds for quantum field
theory at extremely high energies.

\tableofcontents{}
\[
\]

\end{abstract}
\maketitle

\section{Introduction}

The field equations of Einstein's gravity, even in vacuum $R_{\mu\nu}=0$,
are highly nonlinear, but still there is an impressive collection
of exact solutions: some describing spacetimes outside compact sources,
some describing nonlinear waves in curved or flat backgrounds, and
some providing idealized cosmological spacetimes \emph{etc}. According
to the lore in effective field theories, the Einstein-Hilbert action
will be modified, or one might say, quantum-corrected after heavy
degrees of freedom in the microscopic theory are integrated out, with
higher powers curvature and its derivatives at small distances/high
energies. The ensuing theory at a given high energy scale could be
a very complicated one with an action of the form
\begin{equation}
I=\int d^{D}x\sqrt{-g}f\left(g,R,\nabla R,\dots\right),\label{eq:Generic_gravity_theories}
\end{equation}
where $f$ is a smooth function of its arguments, which are the metric
$g$ and the Riemann tensor denoted simply as $R$. Of course, it
is quite possible that there are additionally nonminimally coupled
fields such as scalar fields taking part in gravitation. But, in what
follows we shall assume that this is not the case and gravity is simply
described by the metric. This UV-corrected theory is much more complicated
than Einstein's gravity, and so one might have a priori very little
hope of finding exact solutions. Of course, what is even worse is
that beyond the first few terms in perturbation theory, we do not
really know the form of this modified theory at a given high energy
scale. Hence, apparently, in the absence of the field equations, one
may refrain from searching for solutions, but it turns out that the
situation is not hopeless: there is an interesting line of research
that started some time ago with the works \cite{Gibbons,Deser,Gueven,Amati-Klimcik,Horowitz-Steif,Horowitz-Tseytlin,Coley}
and culminated into a highly fertile research avenue. The idea is
to find metrics, so called \emph{universal metrics} \cite{Hervik},
that solve all the metric-based field equations of quantum-corrected
gravity, with slight modifications in the parameters that reflect
the underlying theory. The notion of universal metrics, with refinements
such as strongly and weakly universal were made in \cite{Hervik},
we shall not go into that distinction here and we shall also not go
into the distinction of critical versus non-critical metrics, where
the former extremize an action while the latter solve a covariantly
conserved field equation not necessarily coming from an action. These
universal metrics, in addition to being valuable on their own, provide
potentially consistent backgrounds for quantum field theory at extremely
high energies where the backreaction or gravity of the quantum fields
cannot be neglected. \textit{\emph{Universal Einsteinian (Ricci-flat
or Einstein space) metrics were studied in the works \cite{Universal-TypeIII-N,Universal-TypeII}.
Non-Einsteinian universal metrics, such as the ones considered here,
with or without cosmological constant are very rare.}}

From the above discussion, it should be clear that finding such universal
metrics is a highly nontrivial task; hence, in the literature, there
does not exist many examples save the ones we quoted above. But, recently,
we have provided new examples of universal metrics: we have shown
that the AdS-plane wave \cite{Gullu-Gurses,fRicci,GravWaves3D} (see
also \cite{Alishah}) and the AdS-spherical wave \cite{gurses1,GravWaves3D}
metrics built on the (anti)-de Sitter {[}(A)dS{]} backgrounds solve
generic gravity theories with an action of the form (\ref{eq:Generic_gravity_theories})
or in general covariant field equations that satisfy a Bianchi identity
\cite{Gurses-PRL,AdS-plane_pp-wave,GravWaves3D,GRF}. These previously
found examples are in the form of the Kerr-Schild metrics%
\footnote{Higher dimensional Kerr-Schild spacetimes are extensively studied
in \cite{Ortaggio,pravda2}.%
} splitting as
\begin{equation}
g_{\mu\nu}=\bar{g}_{\mu\nu}+2V\lambda_{\mu}\lambda_{\nu},\label{eq:AdS-wave_KS}
\end{equation}
where $\bar{g}_{\mu\nu}$ represents the (A)dS spacetime and the $\lambda$
vector satisfies the following four relations
\begin{equation}
\lambda^{\mu}\lambda_{\mu}=0,\qquad\nabla_{\mu}\lambda_{\nu}=\xi_{(\mu}\lambda_{\nu)},\qquad\xi_{\mu}\lambda^{\mu}=0,\qquad\lambda^{\mu}\partial_{\mu}V=0.\label{eq:AdS-wave_prop}
\end{equation}
Observe that a second vector $\xi$ appears whose definition is given
by the second relation, with the symmetrization convention defined
as $2\xi_{(\mu}\lambda_{\nu)}\equiv\xi_{\mu}\lambda_{\nu}+\lambda_{\mu}\xi_{\nu}$.
Note also that the $\lambda$ vector is not a recurrent vector in
general and hence the spacetime does not have the special holonomy
group $\text{Sim}(D-2)$ as was considered to be the case in \cite{Hervik}.
With the second and third relations, the null $\lambda$ vector becomes
nonexpanding, shear-free, and nontwisting; making (\ref{eq:AdS-wave_KS})
a Kundt spacetime; therefore, we shall call this class of metrics
as the Kerr-Schild--Kundt (KSK) class.%
\footnote{The last condition in (\ref{eq:AdS-wave_prop}) is essential in showing
the universality of the metrics although that property is not included
in the definition of KSK metrics.%
}

In this work, we prove\textit{\emph{ that for any metric of the form
(\ref{eq:AdS-wave_KS}) satisfying the conditions (\ref{eq:AdS-wave_prop}),
the }}covariant field equations coming from the variation of (\ref{eq:Generic_gravity_theories})
without any matter fields\textit{\emph{ reduce to an equation linear
in the traceless-Ricci tensor. This is the main purpose of this work.
Once this reduction is achieved, one can have a further reduction
in the field equations into a form that transparently shows that the
solutions of Einstein's gravity and the quadratic curvature gravity
in the KSK class are also solutions of generic gravity theories. The
Einsteinian solutions are the members of the Type N universal spacetimes
studied in \cite{Universal-TypeIII-N}. In addition to these Einsteinian
universal metrics, the solutions of the quadratic curvature gravity
in the KSK class also solve the}} metric-based source-free field equations
of\textit{\emph{ any generic gravity theory, that is these metrics
are }}\emph{non-Einsteinian} universal metrics\textit{\emph{. As we
stated above, the AdS-plane wave and the AdS-spherical wave metrics
belong to the non-Einsteinian KSK family of metrics as being solutions
of the quadratic curvature gravity theories. In addition, rather recently,
we proposed a solution generation technique \cite{GRF} to construct
non-Einsteinian universal metrics and we found a new member of this
class which is the dS-hyperbolic wave metric \cite{dS-waves}.}}

For metrics \textit{\emph{of the form (\ref{eq:AdS-wave_KS}) satisfying
the conditions (\ref{eq:AdS-wave_prop})}}, the vacuum field equations
of the generic gravity theory with the action (\ref{eq:Generic_gravity_theories})
can be written as
\begin{equation}
E_{\mu\nu}\equiv eg_{\mu\nu}+\sum_{n=0}^{N}a_{n}\square^{n}\, S_{\mu\nu}=0,\label{eq:EoM_generic}
\end{equation}
as an immediate consequence of the Theorem 1 to be proven in Section
\ref{sec:Universality-Of-KSK}. Here, $S_{\mu\nu}$ is the traceless-Ricci
tensor, and $\square$ is the d'Alembert operator with respect to
the metric $g_{\mu\nu}$. The derivative order of the generic theory
is $2N+2$ such that $N=0$ is Einstein's gravity (or the Einstein--Gauss-Bonnet
theory) and $N=1$ is the quadratic curvature gravity (or $f\left(g,R\right)$
where $R$ represents the Riemann tensor). The field equations split
into a single trace part and a higher derivative nonlinear wave equation
for the traceless part. Taking the trace of this equation yields a
scalar equation

\noindent
\begin{equation}
e=0,
\end{equation}

\noindent which determines the effective cosmological constant in
terms of the parameters of the theory, such as the bare cosmological
constant and the dimensionful parameters that appear in front of the
curvature invariants. On the other hand, the traceless part is a nontrivial
nonlinear equation
\begin{equation}
\sum_{n=0}^{N}a_{n}\square^{n}S_{\mu\nu}=0.\label{eq:EoM_general}
\end{equation}
This reduction is highly impressive, but in this form, the above equation
cannot be solved save for some trivial cases. Hence, a further reduction
is needed. It was shown in \cite{AdS-plane_pp-wave} that this is
possible as
\begin{equation}
\square^{n}S_{\mu\nu}=\left(-1\right)^{n}\lambda_{\mu}\lambda_{\nu}\left(\mathcal{O}+\frac{2}{\ell^{2}}\right)^{n}\mathcal{O}V.\label{eq:Box^n_Smn}
\end{equation}
Here, the operator $\mathcal{O}$ is defined as\textit{\emph{
\begin{equation}
\mathcal{O}\equiv\square+2\xi^{\mu}\partial_{\mu}+\frac{1}{2}\xi^{\mu}\xi_{\mu}-\frac{2\left(D-2\right)}{\ell^{2}}=\bar{\square}+2\xi^{\mu}\partial_{\mu}+\frac{1}{2}\xi^{\mu}\xi_{\mu}-\frac{2\left(D-2\right)}{\ell^{2}},\label{eq:Operator_defn}
\end{equation}
}}where $\bar{\square}$ is the d'Alembert operator with respect to
the background metrics $\bar{g}_{\mu\nu}$\textit{\emph{ and }}$S_{\mu\nu}=-\lambda_{\mu}\lambda_{\nu}\mathcal{O}V$.
This result given in (\ref{eq:Box^n_Smn}) is valid for the KSK class
with any $\xi_{\mu}$ satisfying (\ref{eq:AdS-wave_prop}), and using
this, (\ref{eq:EoM_general}) reduces to a linear equation
\begin{equation}
\lambda_{\mu}\lambda_{\nu}\sum_{n=0}^{N}a_{n}\left(-1\right)^{n}\left(\mathcal{O}+\frac{2}{\ell^{2}}\right)^{n}\mathcal{O}V=0.\label{eq:EoM_general-1}
\end{equation}

\noindent For $N\ge1$, this equation can be factorized as
\begin{equation}
\prod_{n=0}^{N}\left(\mathcal{O}+b_{n}\right)\,\mathcal{O}\, V=0,\label{eq:EoM_product_form}
\end{equation}
where $b_{n}$ is related to $a_{n}$s and so to the parameters of
the theory; albeit, in general, in a complicated implicit way.\textit{\emph{
If all $b_{n}$s are distinct and none is zero, the most general solution
of (\ref{eq:EoM_product_form}) is in the form
\begin{equation}
\ensuremath{V=V_{E}+V_{1}+V_{2}+\dots+V_{N}},
\end{equation}
where $V_{E}$ is the Einsteinian solution satisfying
\begin{equation}
\mathcal{O}V_{E}=0,\label{eq:Einsteinian_eqn}
\end{equation}
and $V_{n}$ is the solution of the quadratic curvature gravity satisfying
\begin{equation}
\left(\mathcal{O}+b_{n}\right)\, V_{n}=0,\label{eq:Quad_curv_eqn}
\end{equation}
for all $n=1,2,\cdots,N$. For example, when $N=1$, $V=V_{E}+V_{1}$
represents the quadratic curvature gravity solutions which also solve
the generic theory. On the other hand, if some $b_{n}$s coincide
or vanish, then genuinely fourth or higher power operators, such as
$\left(\mathcal{O}+b_{n}\right)^{2}$, arise with Log-type solutions
having asymptotically non-AdS behavior which exist in the so-called
critical theories. Since $\mathcal{O}$ given in (\ref{eq:Operator_defn})
is an operator which solely depends on the background metrics (flat,
AdS, or dS), the solutions of (\ref{eq:Einsteinian_eqn}) and (\ref{eq:Quad_curv_eqn})
for $V_{E}$ and $V_{n}$ can easily be obtained by using some known
techniques such as the method of separation of variables or the method
of Green's function. As we have studied such issues in other works
such as \cite{Gullu-Gurses,gurses1,GravWaves3D}, here we shall not
consider particular cases but give a detailed proof of how KSK metrics
are universal provided that the equations (\ref{eq:Einsteinian_eqn})
and (\ref{eq:Quad_curv_eqn}) are solved for the functions $V_{E}$
and $V_{n}$. In the rest of the paper, we call the KSK metrics where
the metric function $V$ solves (\ref{eq:Einsteinian_eqn})--(\ref{eq:Quad_curv_eqn})
as universal.}}

The layout of the paper is as follows: In Section II, we give the
curvature properties of the KSK metrics as well as the relations satisfied
by the two special vectors $\lambda$ and $\xi$ that are important
in description of these spacetimes. Section III constitutes the bulk
of the paper where we show that the KSK metrics are universal. In
the Appendix, we give an alternative proof by mathematical induction.
As our claim is strong, we were compelled to give two proofs which
can be read independently. The one in the bulk of the paper is shorter
but the one in the Appendix comes with various examples that will
help the reader appreciate the construction.

\section{Curvature tensors and Properties of Kerr-Schild--Kundt class}

In what follows, $D$ will denote the number of dimensions of the
spacetime. The properties of the KSK type metrics were previously
discussed in \cite{gurses1,AdS-plane_pp-wave}. Here, we shall briefly
recapitulate some of these which will be crucial in the proof and
we shall also give some additional constructions in this section.
The scalar curvature of KSK metrics is constant and normalized%
\footnote{Here, the relation between the effective cosmological constant $\Lambda$
and the AdS radius $\ell$ is given as
\[
-\frac{1}{\ell^{2}}\equiv\frac{2\Lambda}{\left(D-1\right)\left(D-2\right)}.
\]
} as $R=-D\left(D-1\right)/\ell^{2}$ and the traceless-Ricci tensor,
$S_{\mu\nu}\equiv R_{\mu\nu}-\frac{R}{D}g_{\mu\nu}$, can be shown
to satisfy
\begin{equation}
S_{\mu\nu}=\rho\lambda_{\mu}\lambda_{\nu},\label{eq:Smn_KSK}
\end{equation}
where \textit{\emph{of course $\lambda_{\mu}$ is the vector appearing
in metric (\ref{eq:AdS-wave_KS}) and the new object $\rho$ is given
in terms of an operator acting on the profile function $V$ as
\begin{equation}
\rho=-\mathcal{O}V=-\left(\square+2\xi^{\mu}\partial_{\mu}+\frac{1}{2}\xi^{\mu}\xi_{\mu}-\frac{2\left(D-2\right)}{\ell^{2}}\right)V.\label{eq:rho_defn}
\end{equation}
}}This expression is not difficult to obtain, but a more involved
computation gives the Weyl tensor as %
\footnote{\textit{\emph{The anti-symmetrization with the square brackets is
weighted with 1/2.}}%
}
\begin{equation}
C_{\mu\alpha\nu\beta}=4\lambda_{[\mu}\Omega_{\alpha][\beta}\lambda_{\nu]},\label{eq:Weyl_KSK}
\end{equation}
\textit{\emph{where the symmetric two-tensor $\Omega_{\alpha\beta}$
is given as
\begin{equation}
\Omega_{\alpha\beta}\equiv-\left[\nabla_{\alpha}\partial_{\beta}+\xi_{(\alpha}\partial_{\beta)}+\frac{1}{2}\xi_{\alpha}\xi_{\beta}-\frac{1}{D-2}g_{\alpha\beta}\left(\mathcal{O}+\frac{2\left(D-2\right)}{\ell^{2}}\right)\right]V,\label{eq:Omega_defn}
\end{equation}
Its contraction with the $\lambda$ vector and its trace read }}

\[
\lambda^{\alpha}\Omega_{\alpha\beta}=\frac{1}{2}\lambda_{\beta}\Omega_{\alpha}^{\alpha},\qquad\Omega_{\alpha}^{\alpha}=\xi^{\alpha}\partial_{\alpha}V-\frac{2}{D-2}\rho+\frac{4}{\ell^{2}}V,
\]
which make it clear that the Weyl tensor satisfies $\lambda^{\mu}C_{\mu\alpha\nu\beta}=0$.
Observe that just like the metric function $V$, due to the Bianchi
identity and the constancy of the scalar curvature, one has $\nabla^{\mu}S_{\mu\nu}=0$
yielding
\begin{equation}
\lambda^{\mu}\nabla_{\mu}\rho=0,
\end{equation}
which also follows from an explicit calculation using the definition
(\ref{eq:rho_defn}) and $\lambda^{\mu}\nabla_{\mu}V=0$.

Let us now calculate the Riemann tensor: using the decomposition
\begin{equation}
R_{\mu\alpha\nu\beta}=C_{\mu\alpha\nu\beta}+\frac{2}{D-2}\left(g_{\mu[\nu}S_{\beta]\alpha}-g_{\alpha[\nu}S_{\beta]\mu}\right)+\frac{2R}{D\left(D-1\right)}g_{\mu[\nu}g_{\beta]\alpha},\label{eq:Riem_in_C_S_gR}
\end{equation}
one arrives at a compact form for the KSK metrics
\begin{equation}
R_{\mu\alpha\nu\beta}=4\lambda_{[\mu}\Theta_{\alpha][\beta}\lambda_{\nu]}+\frac{2R}{D\left(D-1\right)}g_{\mu[\nu}g_{\beta]\alpha},\label{eq:Riem_in_Theta}
\end{equation}
where $\Theta_{\alpha\beta}$ is defined in terms $\rho$ and $\Omega_{\alpha\beta}$
as
\begin{equation}
\Theta_{\alpha\beta}\equiv\Omega_{\alpha\beta}+\frac{1}{D-2}\rho g_{\alpha\beta}=-\left(\nabla_{\alpha}\partial_{\beta}+\xi_{(\alpha}\partial_{\beta)}+\frac{1}{2}\xi_{\alpha}\xi_{\beta}-\frac{2}{\ell^{2}}g_{\alpha\beta}\right)V.\label{eq:Theta_defn}
\end{equation}
We shall make use of this form of the Riemann tensor in the next section.
The trace and $\lambda^{\alpha}$ contraction of the two-tensor $\Theta_{\alpha\beta}$
are
\begin{equation}
\Theta_{\alpha}^{\alpha}=\rho+\xi^{\alpha}\partial_{\alpha}V+\frac{4V}{\ell^{2}},\qquad\lambda^{\alpha}\Theta_{\alpha\beta}=\frac{1}{2}\lambda_{\beta}\Bigl(\Theta_{\alpha}^{\alpha}-\rho\Bigr).
\end{equation}
All of these expressions are exact even though the metric function
$V$ appears linearly, which shows the remarkable property of the
Kerr-Schild metrics in addition to the properties we have listed,
defining the KSK class.

Finally, for the KSK metrics, we need the following identities: once-contracted
Bianchi identity
\begin{equation}
\nabla^{\nu}R_{\mu\alpha\nu\beta}=\nabla_{\mu}R_{\alpha\beta}-\nabla_{\alpha}R_{\mu\beta},
\end{equation}
for constant $R$ yields
\begin{equation}
\nabla^{\nu}R_{\mu\alpha\nu\beta}=\nabla_{\mu}S_{\alpha\beta}-\nabla_{\alpha}S_{\mu\beta},
\end{equation}
which then leads to the double-divergence of the Riemann tensor
\begin{equation}
\nabla^{\mu}\nabla^{\nu}R_{\mu\alpha\nu\beta}=\left(\square-\frac{R}{D-1}\right)S_{\alpha\beta}.\label{eq:Double_derv_Riem_to_S}
\end{equation}
In obtaining this identity, we made use of $\nabla^{\mu}\nabla_{\sigma}S_{\mu\nu}=\frac{R}{D-1}S_{\sigma\nu}$
which follows from
\begin{align}
\nabla^{\mu}\nabla_{\sigma}S_{\mu\nu} & =\left[\nabla_{\mu},\nabla_{\sigma}\right]S_{\nu}^{\mu}=R_{\sigma\alpha}S_{\nu}^{\alpha}+R_{\mu\sigma\nu}^{\phantom{\mu\sigma\nu}\alpha}S_{\alpha}^{\mu},
\end{align}
after using the contractions $R_{\sigma\alpha}S_{\nu}^{\alpha}=\frac{R}{D}S_{\sigma\nu}$
and $R_{\mu\alpha\nu\beta}S^{\mu\beta}=\frac{R}{D\left(D-1\right)}S_{\nu\alpha}$.

The $\xi$ vector that does not appear in the metric but appears in
the definition of the KSK class will play an important role in the
proof below; therefore, let us work out some of the identities that
it satisfies:
\begin{equation}
\lambda^{\nu}\nabla_{\mu}\xi_{\nu}=-\frac{1}{2}\lambda_{\mu}\xi^{\nu}\xi_{\nu},\label{eq:1st_id}
\end{equation}
and its divergence is
\begin{equation}
\nabla_{\mu}\xi^{\mu}=-\frac{1}{4}\xi^{\mu}\xi_{\mu}+\frac{2D-3}{D\left(D-1\right)}R.\label{eq:Div_of_ksi}
\end{equation}
We also have\textit{\emph{
\begin{equation}
\lambda^{\mu}\nabla_{\mu}\xi_{\alpha}=-\lambda_{\alpha}\left(\frac{1}{4}\xi^{\mu}\xi_{\mu}-\frac{1}{D\left(D-1\right)}R\right).\label{eq:Directional_deriv_of_ksi}
\end{equation}
The first equality is simply due to $\lambda^{\nu}\xi_{\nu}=0$. To
obtain the second}}%
\footnote{\textit{\emph{A variation of (\ref{eq:Div_of_ksi}) appeared in the
Appendix B of \cite{gurses1} such that it involves the covariant
derivative with respect to the Christoffel connection of AdS, that
is $\bar{\nabla}_{\mu}$. Thus, another way to obtain (\ref{eq:Div_of_ksi})
is to show the equivalence $\bar{\nabla}_{\mu}\xi^{\mu}=\nabla_{\mu}\xi^{\mu}$.
This result immediately follows from the fact that the Christoffel
connection of the AdS spacetime is related to the Christoffel connection
of the full metric as (see, for example, Appendix B of \cite{gurses1})
\[
\Omega_{\phantom{\mu}\alpha\beta}^{\mu}\equiv\Gamma_{\alpha\beta}^{\mu}-\bar{\Gamma}_{\alpha\beta}^{\mu}=\bar{\nabla}_{\alpha}\left(V\lambda^{\mu}\lambda_{\beta}\right)+\bar{\nabla}_{\beta}\left(V\lambda^{\mu}\lambda_{\alpha}\right)-\bar{\nabla}^{\mu}\left(V\lambda_{\alpha}\lambda_{\beta}\right),
\]
and using the fact that $\Omega_{\phantom{\mu}\mu\beta}^{\mu}=0$,
one has $\bar{\Gamma}_{\mu\beta}^{\mu}=\Gamma_{\mu\beta}^{\mu}$. }}%
}\textit{\emph{ and the third identities, let us note that we have
$\left[\nabla_{\mu},\nabla_{\nu}\right]\lambda_{\beta}=R_{\mu\nu\beta}^{\phantom{\mu\nu\beta}\rho}\lambda_{\rho}$,
whose right-hand side reduces to
\begin{equation}
R_{\mu\nu\beta}^{\phantom{\mu\nu\beta}\rho}\lambda_{\rho}=\frac{R}{D\left(D-1\right)}\left(g_{\mu\beta}\lambda_{\nu}-\lambda_{\mu}g_{\nu\beta}\right),
\end{equation}
after using (\ref{eq:Riem_in_C_S_gR}) and the fact that }}the KSK
spacetime is Type-N Weyl (\ref{eq:Weyl_KSK}) and Type-N traceless-Ricci\textit{\emph{
(\ref{eq:Smn_KSK}) \cite{ClassifyHighDim,ClassifyHighDimRev}. On
the other hand, the left-hand side, $\left[\nabla_{\mu},\nabla_{\nu}\right]\lambda_{\beta}$,
takes the form
\begin{equation}
\left[\nabla_{\mu},\nabla_{\nu}\right]\lambda_{\beta}=\lambda_{[\nu}\nabla_{\mu]}\xi_{\beta}-\lambda_{\beta}\nabla_{[\nu}\xi_{\mu]}-\frac{1}{2}\xi_{\beta}\lambda_{[\nu}\xi_{\mu]},\label{eq:lambda_double_anti-sym_derv}
\end{equation}
after using $\nabla_{\mu}\lambda_{\nu}=\xi_{(\mu}\lambda_{\nu)}$
recursively. Overall, one has
\begin{equation}
2\lambda_{[\nu}\nabla_{\mu]}\xi_{\beta}-2\lambda_{\beta}\nabla_{[\nu}\xi_{\mu]}-\xi_{\beta}\lambda_{[\nu}\xi_{\mu]}=\frac{2R}{D\left(D-1\right)}\left(g_{\mu\beta}\lambda_{\nu}-\lambda_{\mu}g_{\nu\beta}\right),
\end{equation}
which can be used to find $\nabla_{\mu}\xi^{\mu}$ and $\lambda^{\mu}\nabla_{\mu}\xi_{\alpha}$
after performing the $g^{\mu\beta}$ and $\lambda^{\mu}$ contractions
yielding}}%
\textit{\emph{
\begin{equation}
\lambda^{\mu}\nabla_{\mu}\xi_{\nu}=-\lambda_{\nu}\left(\nabla_{\mu}\xi^{\mu}+\frac{1}{2}\xi^{\mu}\xi_{\mu}-\frac{2R}{D}\right),\label{eq:g_contraction}
\end{equation}
\begin{equation}
\lambda_{\nu}\lambda^{\mu}\nabla_{\mu}\xi_{\beta}+\lambda_{\beta}\lambda^{\mu}\nabla_{\mu}\xi_{\nu}=-\lambda_{\beta}\lambda_{\nu}\left(\frac{1}{2}\xi^{\mu}\xi_{\mu}-\frac{2R}{D\left(D-1\right)}\right),\label{eq:lambda_contraction}
\end{equation}
respectively, with the use of (\ref{eq:1st_id}). Then, using (\ref{eq:g_contraction})
in (\ref{eq:lambda_contraction}) yields the equation (\ref{eq:Div_of_ksi})
and making use of that equation in (\ref{eq:g_contraction}) yields
(\ref{eq:Directional_deriv_of_ksi}).}}%
\begin{comment}
\textit{\emph{Having (\ref{eq:g_contraction}) in (\ref{eq:lambda_contraction})
yields
\begin{align*}
-\lambda_{\nu}\lambda_{\beta}\left(\nabla_{\mu}\xi^{\mu}+\frac{1}{2}\xi^{\mu}\xi_{\mu}-\frac{2R}{D}\right)-\lambda_{\beta}\lambda_{\nu}\left(\nabla_{\mu}\xi^{\mu}+\frac{1}{2}\xi^{\mu}\xi_{\mu}-\frac{2R}{D}\right) & =-\lambda_{\beta}\lambda_{\nu}\left(\frac{1}{2}\xi^{\mu}\xi_{\mu}-\frac{2R}{D\left(D-1\right)}\right)\\
\nabla_{\mu}\xi^{\mu}+\frac{1}{2}\xi^{\mu}\xi_{\mu}-\frac{2R}{D} & =\frac{1}{4}\xi^{\mu}\xi_{\mu}-\frac{R}{D\left(D-1\right)}\\
\nabla_{\mu}\xi^{\mu} & =-\frac{1}{4}\xi^{\mu}\xi_{\mu}+\frac{\left(2D-3\right)R}{D\left(D-1\right)}.
\end{align*}
}}

Now, having this result in
\[
\lambda^{\mu}\nabla_{\mu}\xi_{\nu}=-\lambda_{\nu}\left(\nabla_{\mu}\xi^{\mu}+\frac{1}{2}\xi^{\mu}\xi_{\mu}-\frac{2R}{D}\right),
\]
yields
\begin{align*}
\lambda^{\mu}\nabla_{\mu}\xi_{\nu} & =-\lambda_{\nu}\left(-\frac{1}{4}\xi^{\mu}\xi_{\mu}+\frac{\left(2D-3\right)R}{D\left(D-1\right)}+\frac{1}{2}\xi^{\mu}\xi_{\mu}-\frac{2R}{D}\right)\\
\lambda^{\mu}\nabla_{\mu}\xi_{\nu} & =-\lambda_{\nu}\left(\frac{1}{4}\xi^{\mu}\xi_{\mu}-\frac{R}{D\left(D-1\right)}\right).
\end{align*}
\end{comment}

The identities (\ref{eq:1st_id}) and (\ref{eq:Directional_deriv_of_ksi})
play a crucial role in the proof below, because they represent the
fact that all possible contractions of $\nabla_{\mu}\xi_{\nu}$ with
a $\lambda$ vector yields a free-index $\lambda$ vector and a reduction
in the order of the derivative on the $\xi$ vector by one.

The vector $\partial_{\mu}V$ also satisfies similar properties like
$\xi_{\mu}$: for both of these vectors, contraction with $\lambda^{\mu}$
is zero and contractions of $\nabla_{\mu}\partial_{\nu}V$ with a
$\lambda$ vector satisfy
\begin{equation}
\lambda^{\mu}\nabla_{\mu}\partial_{\nu}V=\lambda^{\mu}\nabla_{\nu}\partial_{\mu}V=-\frac{1}{2}\lambda_{\nu}\xi^{\mu}\partial_{\mu}V,\label{eq:lambda_cont_with_1st_order_derv_of_V}
\end{equation}
where again a free-index $\lambda$ vector appears and the order of
the derivative on $\partial_{\mu}V$ reduces by one. With this background
information, we are now ready to state and give the proof of the theorem
in the next session.

\section{Universality Of KSK Metrics\label{sec:Universality-Of-KSK}}

Here, we are going to prove the following theorem:
\begin{quotation}
\noindent \textbf{Theorem 1:} \emph{For the Kerr-Schild metrics
\[
g_{\mu\nu}=\bar{g}_{\mu\nu}+2V\lambda_{\mu}\lambda_{\nu},
\]
with the properties
\[
\lambda^{\mu}\lambda_{\mu}=0,\qquad\nabla_{\mu}\lambda_{\nu}=\xi_{(\mu}\lambda_{\nu)},\qquad\xi_{\mu}\lambda^{\mu}=0,\qquad\lambda^{\mu}\partial_{\mu}V=0,
\]
where $\bar{g}_{\mu\nu}$ is the metric of a space of constant curvature
(AdS or dS), any second rank symmetric tensor constructed from the
Riemann tensor and its covariant derivatives can be written as a linear
combination of $g_{\mu\nu}$, $S_{\mu\nu}$, and higher derivatives
of $S_{\mu\nu}$ in the form $\square^{n}\, S_{\mu\nu}$ where $\square$
represents the d'Alembertian with respect to $g_{\mu\nu}$, that is
\[
E_{\mu\nu}=eg_{\mu\nu}+\sum_{n=0}^{N}a_{n}\square^{n}\, S_{\mu\nu}.
\]
}

\vspace{0.4cm}

\end{quotation}
\textbf{Proof: }The proof of this theorem relies on the observation
that any contraction of the $\lambda$ vector with any tensor composed
of $V$ and its covariant derivatives, $\xi$ and its covariant derivatives
always yields a free-index $\lambda$ vector in each term in the resulting
expression. Thus, in constructing two-tensors out of the contractions
of any number of Riemann tensor and its derivatives, one must keep
track of the number of $\lambda$ vectors.

Let us consider a generic two-tensor \textit{\emph{which is constructed
by any number of Riemann tensors and its covariant derivatives. We}}
represent this two-tensor symbolically as
\begin{equation}
E_{\mu\nu}\equiv\left[R^{n_{0}}\left(\nabla^{n_{1}}R\right)\left(\nabla^{n_{2}}R\right)\dots\left(\nabla^{n_{m}}R\right)\right]_{\mu\nu},\label{eq:Generic_2-ten_in_R}
\end{equation}
where $R$ represents the Riemann tensor, the superscripts represent
the number of terms involved such as $n_{0}$ represents the number
of Riemann tensors without covariant derivatives, and $n_{1}\le n_{2}\le\dots\le n_{m}$
is assumed without loss of generality. In the notation of this section,
the Riemann tensor given in (\ref{eq:Riem_in_Theta}) can be simply
given as $R=\lambda^{2}\Theta+g^{2}$. In the above expression, we
omitted the metric tensors among the terms, and in principle, any
contraction pattern is possible. The presence of these metric tensors
does not alter any of our discussions below. It is obvious that to
have a two-tensor, the sum $\sum_{i=1}^{m}n_{i}$ should be even.
Considering the metric compatibility condition and using the form
of the Riemann tensor in (\ref{eq:Riem_in_Theta}), $E_{\mu\nu}$
reduces to (say a new tensor $\mathcal{E}_{\mu\nu}$)
\begin{equation}
\mathcal{E}_{\mu\nu}\equiv\left[\lambda^{2n_{0}}\Theta^{n_{0}}\left(\nabla^{n_{1}}\left[\lambda^{2}\Theta\right]\right)\left(\nabla^{n_{2}}\left[\lambda^{2}\Theta\right]\right)\dots\left(\nabla^{n_{m}}\left[\lambda^{2}\Theta\right]\right)\right]_{\mu\nu},\label{eq:Generic_2-ten_in_theta}
\end{equation}
where we omitted the metrics coming out of the Riemann tensors $R^{n_{0}}$,
since considering them just yields a sum of two-tensor forms updated
with $\lambda^{2n_{r}}\Theta^{n_{r}}$ instead of $\lambda^{2n_{0}}\Theta^{n_{0}}$
where $n_{r}<n_{0}$ always, so these terms are genuinely covered
in $\mathcal{E}_{\mu\nu}$.

Now, let us consider the tensorial structures appearing in $\mathcal{E}_{\mu\nu}$.
First, note that $\Theta$ defined in (\ref{eq:Theta_defn}) is composed
of $V$ and its first and second order derivatives in addition to
the $\xi$ vector. Secondly, let us consider the highest order derivative
term $\left(\nabla^{n_{m}}\left[\lambda^{2}\Theta\right]\right)$
which is a $\left(0,n_{m}+4\right)$ rank tensor. Note that with each
application of the covariant derivative on $\lambda$, one can use
$\nabla_{\mu}\lambda_{\nu}=\xi_{(\mu}\lambda_{\nu)}$; and therefore,
$\left(\nabla^{n_{m}}\left[\lambda^{2}\Theta\right]\right)$ represents
a sum of $\left(0,n_{m}+4\right)$ rank tensors that are built with
$V$ and its up to $\left(n_{m}+2\right)^{{\rm th}}$-order derivatives
in addition to the $\xi$ vector and its $n_{m}^{{\rm th}}$-order
derivatives. Therefore, the $\left(0,s\equiv4n_{0}+4m+\sum_{i=1}^{m}n_{i}\right)$
rank tensor,
\begin{equation}
\mathcal{E}_{\mu_{1}\dots\mu_{s}}\equiv\left[\lambda^{2n_{0}}\Theta^{n_{0}}\left(\nabla^{n_{1}}\left[\lambda^{2}\Theta\right]\right)\left(\nabla^{n_{2}}\left[\lambda^{2}\Theta\right]\right)\dots\left(\nabla^{n_{m}}\left[\lambda^{2}\Theta\right]\right)\right],\label{eq:Generic_tensor_structure}
\end{equation}
represents a sum of $\left(0,s\right)$ rank tensors which are built
with $2\left(n_{0}+m\right)$ number of $\lambda$ vectors and the
remaining $\left(0,s-2n_{0}-2m\right)$ rank tensorial parts are built
with $V$ and its up to $\left(n_{m}+2\right)^{{\rm th}}$-order derivatives
in addition to the $\xi$ vector and its $n_{m}^{{\rm th}}$-order
derivatives.

After discussing the tensorial structure of $\mathcal{E}_{\mu_{1}\dots\mu_{s}}$,
now let us analyze the nature of the $\left(s/2-1\right)$ number
of contractions with the inverse metric yielding $\mathcal{E}_{\mu\nu}$.
First, note that the contractions of the $\lambda^{\mu}$ vector with
$\lambda_{\mu}$, $\xi_{\mu}$, and $\partial_{\mu}V$ yield zero.
Secondly, the contractions of the $\lambda^{\mu}$ vector with the
first order derivatives of $\xi_{\mu}$ and $\partial_{\mu}V$ yield
(\ref{eq:1st_id}) and (\ref{eq:Directional_deriv_of_ksi}), and (\ref{eq:lambda_cont_with_1st_order_derv_of_V}),
respectively. In these contractions, the important points to observe
are:
\begin{itemize}
\item the number of the $\lambda$ vectors is preserved since a free-index
$\lambda$ always appears in the results,
\item contraction with the $\lambda$ vector removes the first order derivatives
acting on $\xi_{\mu}$ and $\partial_{\mu}V$.
\end{itemize}
Now, let us analyze the $\lambda^{\mu}$ contraction of the terms
involving higher order covariant derivatives acting on $\xi_{\mu}$
and $\partial_{\mu}V$. Note that to arrive at the stated proof, instead
of explicit formulae, the tensorial structure of the expressions after
the $\lambda^{\mu}$ contractions is important. Since the $\lambda^{\mu}$
contractions of both $\xi_{\mu}$ and $\partial_{\mu}V$ yield the
same structure, we worked with $\xi_{\mu}$ for definiteness; however,
the conclusions we obtained are also valid in the $\partial_{\mu}V$
case. Thus, let us consider the $\left(0,r+1\right)$ rank tensor
in the form
\begin{equation}
\nabla_{\mu_{1}}\nabla_{\mu_{2}}\dots\nabla_{\mu_{r}}\xi_{\mu_{r+1}}.
\end{equation}
The $\lambda^{\mu}$ contraction can be through one of the covariant
derivatives as
\begin{equation}
\lambda^{\mu}\nabla_{\mu_{1}}\nabla_{\mu_{2}}\dots\nabla_{\mu}\dots\nabla_{\mu_{r-1}}\xi_{\mu_{r}},\label{eq:lambda_derv_cont}
\end{equation}
or through the $\xi$ vector as
\begin{equation}
\lambda^{\mu}\nabla_{\mu_{1}}\nabla_{\mu_{2}}\dots\nabla_{\mu_{r}}\xi_{\mu}.\label{eq:lambda_xi_cont}
\end{equation}
For these two contraction patterns, the tensorial structure of the
final results are sums of the $\left(0,r\right)$ rank tensors satisfying
the properties;
\begin{itemize}
\item each term involves a free-index $\lambda$ vector,
\item for all the terms, the highest order of derivative acting on $\xi$
will be $r-1$ or less.
\end{itemize}
To show these properties, we need to use the basic identities (\ref{eq:1st_id})
and (\ref{eq:Directional_deriv_of_ksi}), and to make such a use,
first, one needs to change the orders of the derivatives in (\ref{eq:lambda_derv_cont})
such that one has
\begin{equation}
\lambda^{\mu}\nabla_{\mu_{1}}\nabla_{\mu_{2}}\dots\nabla_{\mu_{r-1}}\nabla_{\mu}\xi_{\mu_{r}},
\end{equation}
by using the Ricci identity%
\footnote{Here, with Ricci identity, we mean
\[
\left[\nabla_{\mu},\nabla_{\nu}\right]T_{\alpha\beta\dots\gamma}=R_{\mu\nu\alpha}^{\phantom{\mu\nu\alpha}\lambda}T_{\lambda\beta\dots\gamma}+\dots+R_{\mu\nu\gamma}^{\phantom{\mu\nu\gamma}\lambda}T_{\alpha\beta\dots\lambda}.
\]
} producing Riemann tensors for each change of order. After making
all the change of orders and applying simply the product rule for
the covariant derivatives, one arrives at
\begin{equation}
\lambda^{\mu}\nabla_{\mu_{1}}\nabla_{\mu_{2}}\dots\nabla_{\mu}\dots\nabla_{\mu_{r-1}}\xi_{\mu_{r}}=\lambda^{\mu}\nabla_{\mu_{1}}\nabla_{\mu_{2}}\dots\nabla_{\mu_{r-1}}\nabla_{\mu}\xi_{\mu_{r}}+\sum_{p}\lambda^{\mu}\left(\nabla^{p}R_{\mu}\right)\left(\nabla^{r-p-2}\xi\right),\label{eq:Order_change_structure}
\end{equation}
where in the last sum, the $\lambda^{\mu}\left(\nabla^{p}R_{\mu}\right)$
term represents $p$ number of covariant derivatives acting on the
Riemann tensor and one index of the Riemann tensor should be contracted
with $\lambda^{\mu}$. Here, $p$ can have various values depending
on the position of the contracted covariant derivative in (\ref{eq:lambda_derv_cont})
and it can be as small as $0$ and as large as $\left(r-2\right)$.
Once we consider the Riemann tensor $R$ symbolically as $\lambda^{2}\Theta$,
then
\begin{equation}
\lambda^{\mu}\left(\nabla^{p}R_{\mu}\right)=\lambda^{\mu}\left(\nabla^{p}\left[\lambda^{2}\Theta\right]_{\mu}\right),
\end{equation}
represents a sum of terms involving two free-index $\lambda$ vectors
and the remaining $\left(0,p+1\right)$-rank tensor structure is built
with the $\xi$, $\partial V$ vectors, and their covariant derivatives.
In each term in this summation, one higher order covariant derivative
term involving $\xi$ or $\partial V$ must have a $\lambda^{\mu}$
contraction. The derivative order of this $\lambda^{\mu}$ contracted
term is at most $\left(r-1\right)$ for the $\partial V$ vector and
$\left(r-2\right)$ for the $\xi$ vector. This is because $\Theta$
involves the first derivative of the $\partial V$ vector and just
the $\xi$ vector itself, and $p$ can take the maximum value of $\left(r-2\right)$.
To summarize, for the last sum in (\ref{eq:Order_change_structure}),
the properties of the tensorial structure of each term is:
\begin{itemize}
\item there are three $\lambda$ vectors one of which is in the contracted
form and the others are free,
\item the total number of derivatives in these terms is at most $\left(r-1\right)$
for $\partial V$ and $\left(r-2\right)$ for $\xi$, so the order
of the derivative is reduced by 1.
\end{itemize}
So, for these terms, we achieved to show the aimed two properties.

Now, let us focus on the first term in (\ref{eq:Order_change_structure})
and (\ref{eq:lambda_xi_cont}). For these terms, we need to change
the order of the covariant derivatives and the $\lambda^{\mu}$ vector
such that in the end we obtain
\begin{equation}
\nabla_{\mu_{1}}\nabla_{\mu_{2}}\dots\nabla_{\mu_{r-1}}\left(\lambda^{\mu}\nabla_{\mu}\xi_{\mu_{r}}\right),
\end{equation}
\begin{equation}
\nabla_{\mu_{1}}\nabla_{\mu_{2}}\dots\nabla_{\mu_{r-1}}\left(\lambda^{\mu}\nabla_{\mu_{r}}\xi_{\mu}\right),
\end{equation}
respectively, and we can apply the identities (\ref{eq:Directional_deriv_of_ksi})
and (\ref{eq:1st_id}) in these terms. To show how we carry out this
simple change of orders, we consider the first term in (\ref{eq:Order_change_structure})
and the same steps apply for (\ref{eq:lambda_xi_cont}). In commuting
the $\lambda^{\mu}$ vector and the covariant derivatives, we simply
have
\begin{equation}
\lambda^{\mu}\nabla_{\mu_{1}}\nabla_{\mu_{2}}\dots\nabla_{\mu_{r-1}}\nabla_{\mu}\xi_{\mu_{r}}=\nabla_{\mu_{1}}\left(\lambda^{\mu}\nabla_{\mu_{2}}\dots\nabla_{\mu_{r-1}}\nabla_{\mu}\xi_{\mu_{r}}\right)-\left(\nabla_{\mu_{1}}\lambda^{\mu}\right)\nabla_{\mu_{2}}\dots\nabla_{\mu_{r-1}}\nabla_{\mu}\xi_{\mu_{r}},
\end{equation}
where in the second term on the right-hand side, one can apply the
defining property of the $\xi$ vector $\nabla_{\mu}\lambda_{\nu}=\xi_{(\mu}\lambda_{\nu)}$
which reduces the derivative order and introduces a free-index $\lambda$
vector. Then, one has
\begin{align}
\lambda^{\mu}\nabla_{\mu_{1}}\nabla_{\mu_{2}}\dots\nabla_{\mu_{r-1}}\nabla_{\mu}\xi_{\mu_{r}}= & \nabla_{\mu_{1}}\left(\lambda^{\mu}\nabla_{\mu_{2}}\dots\nabla_{\mu_{r-1}}\nabla_{\mu}\xi_{\mu_{r}}\right)\nonumber \\
 & -\frac{1}{2}\xi_{\mu_{1}}\lambda^{\mu}\nabla_{\mu_{2}}\dots\nabla_{\mu_{r-1}}\nabla_{\mu}\xi_{\mu_{r}}\nonumber \\
 & -\frac{1}{2}\lambda_{\mu_{1}}\xi^{\mu}\nabla_{\mu_{2}}\dots\nabla_{\mu_{r-1}}\nabla_{\mu}\xi_{\mu_{r}},\label{eq:First_order_change}
\end{align}
where for the last term, we achieved our aim that
\begin{itemize}
\item a free-index $\lambda$ vector is introduced,
\item the derivative order on $\xi_{\mu_{r}}$ is reduced by one.
\end{itemize}
On the other hand, the second term in (\ref{eq:First_order_change})
still involves a $\lambda^{\mu}$ contraction; but this time, the
order of the derivative acting on $\xi_{\mu_{r}}$ is $\left(r-1\right)$.
For this term, one needs to repeat this ongoing process for the generic
$r^{{\rm th}}$-derivative term. For the next step of the change of
orders, we consider the first term on the right-hand side (\ref{eq:First_order_change})
and change the order of $\lambda^{\mu}$ and $\nabla_{\mu_{2}}$ as
\begin{align}
\lambda^{\mu}\nabla_{\mu_{1}}\nabla_{\mu_{2}}\dots\nabla_{\mu_{r-1}}\nabla_{\mu}\xi_{\mu_{r}}= & \nabla_{\mu_{1}}\nabla_{\mu_{2}}\left(\lambda^{\mu}\nabla_{\mu_{3}}\dots\nabla_{\mu_{r-1}}\nabla_{\mu}\xi_{\mu_{r}}\right)\nonumber \\
 & -\left(\nabla_{\mu_{1}}\nabla_{\mu_{2}}\lambda^{\mu}\right)\left(\nabla_{\mu_{3}}\dots\nabla_{\mu_{r-1}}\nabla_{\mu}\xi_{\mu_{r}}\right)\nonumber \\
 & -\left(\nabla_{\mu_{2}}\lambda^{\mu}\right)\nabla_{\mu_{1}}\nabla_{\mu_{3}}\dots\nabla_{\mu_{r-1}}\nabla_{\mu}\xi_{\mu_{r}}\nonumber \\
 & -\frac{1}{2}\xi_{\mu_{1}}\lambda^{\mu}\nabla_{\mu_{2}}\dots\nabla_{\mu_{r-1}}\nabla_{\mu}\xi_{\mu_{r}}\nonumber \\
 & -\frac{1}{2}\lambda_{\mu_{1}}\xi^{\mu}\nabla_{\mu_{2}}\dots\nabla_{\mu_{r-1}}\nabla_{\mu}\xi_{\mu_{r}}.
\end{align}
Here, again using $\nabla_{\mu}\lambda_{\nu}=\xi_{(\mu}\lambda_{\nu)}$
in the second and third terms yield either $\lambda^{\mu}$ contracted
terms having less number of derivatives than $r$ acting on $\xi$
or terms involving a free-index $\lambda$ vector. Again for the terms
involving the $\lambda^{\mu}$ contraction this ongoing procedure
can be repeated. Thus, one can continue changing the order of the
$\lambda^{\mu}$ vector and the covariant derivatives in the first
term until one arrives at
\begin{equation}
\nabla_{\mu_{1}}\nabla_{\mu_{2}}\dots\nabla_{\mu_{r-1}}\left(\lambda^{\mu}\nabla_{\mu}\xi_{\mu_{r}}\right),
\end{equation}
which reduces to
\begin{equation}
\nabla_{\mu_{1}}\nabla_{\mu_{2}}\dots\nabla_{\mu_{r-1}}\left[-\lambda_{\mu_{r}}\left(\frac{1}{4}\xi^{\mu}\xi_{\mu}-\frac{1}{D\left(D-1\right)}R\right)\right],
\end{equation}
after making use of (\ref{eq:Directional_deriv_of_ksi}). This term
after the use of $\nabla_{\mu}\lambda_{\nu}=\xi_{(\mu}\lambda_{\nu)}$
yields a sum of terms involving a free-index $\lambda$ vector, and
for each term, the derivative order on the $\xi$ vectors are always
less then $r$. With these considerations, the expression in (\ref{eq:lambda_derv_cont})
turns into a sum in which each term either involves a free-index $\lambda$
vector or a $\lambda^{\mu}$ contraction. But, for these terms, the
order of covariant derivatives acting on the $\xi$ vector is always
less than $r$. For the latter kind of terms, one can repeat this
ongoing procedure until to the point of only having terms involving
a free-index $\lambda$ vector, and so no $\lambda^{\mu}$ contractions.
The procedure that we discussed for (\ref{eq:lambda_derv_cont}) can
be applicable to the (\ref{eq:lambda_xi_cont}) contraction pattern
for which the only change will be the application of (\ref{eq:1st_id})
instead of (\ref{eq:Directional_deriv_of_ksi}). Similarly, the analysis
of a generic term involving the $r^{{\rm th}}$ order covariant derivatives
acting on $\partial_{\mu}V$ instead of $\xi_{\mu}$ is exactly the
same, as was noted before.

As a result, the $\lambda^{\mu}$ contraction of a generic term involving
the $r^{{\rm th}}$-order covariant derivative of either the $\xi$
vector or the $\partial V$ vector turns into a sum involving terms
satisfying:
\begin{itemize}
\item each term involves a free-index $\lambda$ vector,
\item in each term, the derivative order acting on $\xi$ or $\partial V$
vectors is always less than $r$.
\end{itemize}
These were the aimed properties.

With this result, let us discuss the contractions in $E_{\mu\nu}$
or more explicitly,

\begin{equation}
E_{\mu\nu}=\left[\left(g^{-1}\right)^{s-1}\mathcal{E}_{\mu_{1}\dots\mu_{s}}\right]_{\mu\nu},
\end{equation}
where $g^{-1}$ represents the inverse metric. It is clear that any
nonzero contraction of $2\left(n_{0}+m\right)$ number of $\lambda$
vectors in (\ref{eq:Generic_tensor_structure}) with the other tensorial
parts involving derivatives of $\xi$ and $\partial V$ vectors always
produces a free-index $\lambda$ vector and reduces the derivative
order. Thus, after every nonzero $\lambda$ contraction, the number
of free-index $\lambda$ vectors is \emph{preserved} as $2\left(n_{0}+m\right)$.
Obviously, one cannot avoid having a nonzero contraction once one
reduces the $\left(0,s\right)$-rank tensor $\mathcal{E}_{\mu_{1}\dots\mu_{s}}$
to a $\left(0,2n_{0}+2m\right)$-rank tensor, whose free indices are
only on the $\lambda$ vectors, and $E_{\mu\nu}$ takes the form
\begin{equation}
E_{\mu\nu}=\left[\left(g^{-1}\right)^{n_{0}+m-1}\lambda^{2\left(n_{0}+m\right)}\right]_{\mu\nu},
\end{equation}
which is zero for $n_{0}+m>1$. After this observation, the only remaining
possibility of having a nonzero two-tensor out of $\mathcal{E}_{\mu_{1}\dots\mu_{s}}$
is to have only two $\lambda$ one-forms from the \emph{outset}, so
either $n_{0}=1$ or $m=1$, implying the presence of only one Riemann
tensor in $\mathcal{E}_{\mu_{1}\dots\mu_{s}}$. Thus, the generic
forms of a \emph{nonzero} two-tensor are
\begin{equation}
\left[R\right]_{\mu\nu},\qquad\left[\nabla^{n}R\right]_{\mu\nu},
\end{equation}
where $n$ is even and $\left[R\right]_{\mu\nu}$ just represents
the Ricci tensor while the second term represents a two-tensor contraction
of
\begin{equation}
\nabla_{\mu_{1}}\nabla_{\mu_{2}}\dots\nabla_{\mu_{n}}R_{\nu_{1}\nu_{2}\nu_{3}\nu_{4}}.\label{eq:nth_derv_Riem}
\end{equation}
In analyzing two-tensor contractions of (\ref{eq:nth_derv_Riem}),
the important observation is that in the process of obtaining a nonzero
two-tensor, one can freely change the order of the covariant derivatives
by using the Ricci identity since all the additional terms involving
a second Riemann tensor just yield a zero at the two-tensor level
as we just proved.%
\footnote{Note that for an order change involving the first two derivatives,
there is a possibility of having an additional nonzero term in the
form $\left[\nabla^{n-2}R\right]_{\mu\nu}$ due to the metric part
in the Riemann tensor (\ref{eq:Riem_in_C_S_gR}).%
} In obtaining a nonzero two-tensor out of (\ref{eq:nth_derv_Riem}),
one can have two contraction possibilities either
\begin{equation}
\nabla_{\mu_{1}}\dots\nabla^{\nu_{1}}\dots\nabla^{\nu_{3}}\dots\nabla_{\mu_{n}}R_{\nu_{1}\nu_{2}\nu_{3}\nu_{4}},\label{eq:1st_cont_of_nth_derv_Riem}
\end{equation}
or
\begin{equation}
\nabla_{\mu_{1}}\nabla_{\mu_{2}}\dots\nabla_{\mu_{n}}R_{\nu_{1}\nu_{2}}.\label{eq:2nd_cont_of_nth_derv_Riem}
\end{equation}
For both of them, the following contractions of the covariant derivatives
are among themselves. Because $\nabla_{\mu_{1}}\nabla_{\mu_{2}}\dots\nabla_{\mu_{n}}R$
is zero as the Ricci scalar $R$ is constant and
\begin{equation}
\nabla_{\mu_{1}}\dots\nabla^{\nu_{1}}\dots\nabla_{\mu_{n}}R_{\nu_{1}\nu_{2}},
\end{equation}
yields a zero since one can change the orders of covariant derivatives
until one obtains $\nabla_{\mu_{1}}\dots\nabla_{\mu_{n}}\nabla_{\nu_{2}}R$.
In (\ref{eq:1st_cont_of_nth_derv_Riem}), one can change the order
of derivatives by the Ricci identity to obtain
\begin{equation}
\nabla_{\mu_{1}}\dots\nabla_{\mu_{n}}\nabla^{\nu_{1}}\nabla^{\nu_{3}}R_{\nu_{1}\nu_{2}\nu_{3}\nu_{4}}=\nabla_{\mu_{1}}\dots\nabla_{\mu_{n}}\left(\square-\frac{R}{D-1}\right)S_{\nu_{2}\nu_{4}},\label{eq:1st_S_derv_2ten_form}
\end{equation}
where we used (\ref{eq:Double_derv_Riem_to_S}). Note also that (\ref{eq:2nd_cont_of_nth_derv_Riem})
becomes
\begin{equation}
\nabla_{\mu_{1}}\nabla_{\mu_{2}}\dots\nabla_{\mu_{n}}R_{\nu_{1}\nu_{2}}=\nabla_{\mu_{1}}\nabla_{\mu_{2}}\dots\nabla_{\mu_{n}}S_{\nu_{1}\nu_{2}}.\label{eq:2nd_S_derv_2ten_form}
\end{equation}
The remaining free-indices in the covariant derivatives of (\ref{eq:1st_S_derv_2ten_form})
and (\ref{eq:2nd_S_derv_2ten_form}) can be rearranged such that one
has
\begin{equation}
\square^{\frac{n-2}{2}}\left(\square-\frac{R}{D-1}\right)S_{\nu_{2}\nu_{4}},
\end{equation}
and $\square^{n/2}S_{\nu_{1}\nu_{2}}$, respectively. Note that for
a change of order involving the first two derivatives, it may seem
that there is a possibility of having additional nonzero terms due
to the metric part in (\ref{eq:Riem_in_C_S_gR}). But, one never needs
such a change since for a term in the form
\begin{equation}
\nabla^{\mu_{1}}\nabla_{\mu_{2}}\dots\nabla_{\mu_{1}}\dots\nabla_{\mu_{n}}S_{\nu_{1}\nu_{2}},
\end{equation}
one may only move $\nabla_{\mu_{1}}$ to obtain
\begin{equation}
\square\nabla_{\mu_{2}}\dots\nabla_{\mu_{n}}S_{\nu_{1}\nu_{2}}.
\end{equation}
As a result, the generic two-tensor $E_{\mu\nu}$ \textit{\emph{constructed
from any number of Riemann tensors and its covariant derivatives}}
can be written as a sum of the metric, $S_{\mu\nu}$, and \textit{\emph{higher
derivatives of $S_{\mu\nu}$ in the form $\square^{n}\, S_{\mu\nu}$.
This proves the theorem.}}\emph{\QEDA}\textit{\emph{ }}

\textit{\emph{In the Appendix, we give another, mathematical induction
based, proof of the theorem.}}

\section{Conclusions}

We have shown that the Kerr-Schild--Kundt class of metrics, defined
by the relations (\ref{eq:AdS-wave_KS}) and (\ref{eq:AdS-wave_prop}),
are universal in the sense that they solve the most general quantum-corrected
source-free gravity equations based on the metric tensor, the Riemann
tensor and its arbitrary number of covariant derivatives and their
powers. Our proof here boils down to showing that the generic two-tensor
built out of the contractions of the Riemann tensor and its covariant
derivatives can be written as a symmetric, covariantly-conserved,
two-tensor $E_{\mu\nu}$ for the KSK-class in the form
\begin{equation}
E_{\mu\nu}=eg_{\mu\nu}+\sum_{n=0}^{N}a_{n}\square^{n}\, S_{\mu\nu},
\end{equation}
where $e$ and $a_{n}$ are parameters, constants, of the theory.
One further reduction gives the product of scalar wave type equations
(\ref{eq:EoM_product_form}), generically one of them is massless
and the rest are massive. The massless one corresponds to the Einstein's
theory, and the massive ones correspond to quadratic gravity. Of course,
one must still solve these equations to actually find explicit solutions:
namely, one must determine the metric function $V$. We have not done
this in the current work because, earlier, we already gave examples
of these metrics such as the AdS-plane and AdS-spherical waves as
solutions to quadratic and generic gravity theories \cite{Gullu-Gurses,gurses1,GravWaves3D}.
In \cite{GRF,dS-waves}, we give a systematic way of constructing
solutions, such universal metrics, from curves living in one less
dimension and extend the discussion to the de Sitter case.

\section{Acknowledgments}

This work is partially supported by TUBITAK. M.~G. and B.~T. are
supported by the TUBITAK grant 113F155. T.~C.~S. is supported by
the Science Academy\textquoteright s Young Scientist Awards Program
(BAGEP 2015). T.~C.~S. thanks the Centro de Estudios Cientificos
(CECs) where part of this work was carried out under the support of
Fondecyt with grant 3140127. We would like to thank M.~Ortaggio for
his constructive comments.

\appendix
%dummy comment inserted by tex2lyx to ensure that this paragraph is not empty

\section{Alternative Proof by Induction}

In this Appendix, for a second proof alternative to the one given
in the bulk of the paper, we give necessary recursion relations satisfied
by the tensors in KSK spacetimes. A generic two-tensor constructed
out of the Riemann tensor and its covariant derivatives can be represented
as
\begin{equation}
E_{\mu\nu}\equiv\left[R^{n_{0}}\left(\nabla^{n_{1}}R\right)\left(\nabla^{n_{2}}R\right)\dots\left(\nabla^{n_{m}}R\right)\right]_{\mu\nu},\label{eq:Generic_2-ten_in_R-1}
\end{equation}
where the Riemann tensor $R$ for KSK metrics is
\begin{equation}
R_{\mu\alpha\nu\beta}=4\lambda_{[\mu}\Theta_{\alpha][\beta}\lambda_{\nu]}+\frac{4\Lambda}{\left(D-1\right)\left(D-2\right)}g_{\mu[\nu}g_{\beta]\alpha}.\label{eq:Riem_in_Theta-1}
\end{equation}
Here, $\Theta_{\alpha\beta}$ is defined as
\begin{equation}
\Theta_{\alpha\beta}=-\left(\nabla_{\alpha}\partial_{\beta}+\xi_{(\alpha}\partial_{\beta)}+\frac{1}{2}\xi_{\alpha}\xi_{\beta}-\frac{2}{\ell^{2}}g_{\alpha\beta}\right)V.\label{eq:Theta_defn-1}
\end{equation}
Assuming $n_{m}$ is to be the largest integer, the $\left(0,s\equiv4n_{0}+4m+\sum_{n_{i}=1}^{i=m}n_{i}\right)$
rank tensor,
\begin{equation}
\mathcal{E}_{\mu_{1}\dots\mu_{s}}\equiv\left[R^{n_{0}}\left(\nabla^{n_{1}}R\right)\left(\nabla^{n_{2}}R\right)\dots\left(\nabla^{n_{m}}R\right)\right],\label{eq:Generic_tensor_structure-1}
\end{equation}
represents a sum of rank $\left(0,s\right)$ tensors which can be
decomposed into $2\left(n_{0}+m\right)$ number of $\lambda$ vectors
and rank $\left(0,s-2n_{0}-2m\right)$ tensor structures which are
built of the contractions of the following building blocks\textit{\emph{
\begin{equation}
g_{\mu_{1}\mu_{2}},\qquad\xi_{\mu_{1}},\qquad\left(\prod_{i=1}^{r}\nabla_{\mu_{i}}\right)\xi_{\mu_{r+1}},\qquad\left(\prod_{i=1}^{r+2}\nabla_{\mu_{i}}\right)V,\qquad r=1,2,\dots,n_{m}.
\end{equation}
We need to understand the contractions of $\lambda$ with these building
blocks. For this purpose, we need the following definitions:}}

\textbf{\vspace{0.5cm}
}

\noindent \textbf{Definition 1 -- $\lambda$-reducible tensor:}\textit{
}\textit{\emph{A tensor $E$ of rank $\left(0,m\right)$ is called
$\lambda$-reducible if it can be written as
\[
E_{\mu_{1}\mu_{2}\cdots\mu_{m}}=\sum_{s=1}^{m}\lambda_{\mu_{s}}F_{\mu_{r_{_{1}}}\mu_{r_{_{2}}}\dots\mu_{r_{_{m-1}}}}^{\left(s\right)},
\]
where $\left(r_{1},r_{2},\dots,r_{m-1}\right)$ is an }}\textit{increasing}\textit{\emph{
sequence constructed with the elements of $\left\{ 1,2,\dots,m\right\} \backslash\left\{ s\right\} $
(where the notation $\backslash$ denotes the set-theoretic difference,
that is $s$ is omitted from the set), }}and $F^{\left(s\right)}$
tensors are rank $\left(0,m-1\right)$ tensors containing no free
$\lambda$ vectors. \textbf{\vspace{0.5cm}
}

\noindent \textbf{Definition 2 -- $\lambda$-weight of a tensor: }A
tensor $E$ of rank $\left(0,m\right)$ has the $\lambda$-weight
$n$ if it can be written as a linear combination of $\left(0,m\right)$
rank tensors which can be decomposed into $n$ number of $\lambda$
vectors and rank $\left(0,m-n\right)$ tensors $F^{\left(s\right)}$
which are \emph{not} $\lambda$-reducible, that is
\begin{equation}
E_{\mu_{1}\mu_{2}\cdots\mu_{m}}=\sum_{s=1}^{N}\lambda_{\mu_{_{k_{_{1}}^{s}}}}\lambda_{\mu_{_{k_{_{2}}^{s}}}}\dots\lambda_{\mu_{_{k_{_{n}}^{s}}}}F_{\mu_{r_{_{1}}}\mu_{r_{_{2}}}\dots\mu_{r_{_{m-n}}}}^{\left(s\right)},
\end{equation}
\textit{\emph{where $N$ is the number of the $n$-element subsets
of $\left\{ 1,2,\dots,m\right\} $, that is $N=\left(\begin{array}{c}
m\\
n
\end{array}\right)$, $s$ is the label for each of these $n$-element subsets such that
for each $s$, $\left\{ k_{1}^{s},k_{2}^{s},\dots,k_{n}^{s}\right\} $
is one of these $n$-element subsets, and $\left(r_{1},r_{2},\dots,r_{m-n}\right)$
is an }}\textit{increasing}\textit{\emph{ sequence constructed with
the elements of $\left\{ 1,2,\dots,m\right\} \backslash\left\{ k_{1}^{s},k_{2}^{s},\dots,k_{n}^{s}\right\} $.}}

\textbf{\vspace{0.2cm}
}

\noindent \textbf{Remark:} All the $F$ tensors in the following discussions
are assumed to be \emph{not} $\lambda$-reducible. \textbf{\vspace{0.2cm}
}

\noindent \textbf{Example 1:} The Weyl tensor $C_{\mu\alpha\nu\beta}$
has the $\lambda$-weight 2 since it reads for the KSK class as
\begin{equation}
C_{\mu\alpha\nu\beta}=\lambda_{\mu}\lambda_{\nu}\,\Omega_{\alpha\beta}+\lambda_{\alpha}\lambda_{\beta}\,\Omega_{\mu\nu}-\lambda_{\mu}\lambda_{\beta}\,\Omega_{\alpha\nu}-\lambda_{\alpha}\lambda_{\nu}\,\Omega_{\mu\beta},
\end{equation}
or
\begin{equation}
C_{\mu_{1}\mu_{2}\mu_{3}\mu_{4}}=\sum_{s=1}^{6}\lambda_{\mu_{k_{_{1}}^{s}}}\lambda_{\mu_{k_{_{2}}^{s}}}F_{\mu_{r_{_{1}}}\mu_{r_{_{2}}}}^{\left(s\right)},
\end{equation}
where for the subsets $\left\{ 1,2\right\} $ and $\left\{ 3,4\right\} $,
\textit{\emph{$F_{\mu_{r_{_{1}}}\mu_{r_{_{2}}}}^{\left(s\right)}=0$,
while for the others, $F_{\mu_{r_{_{1}}}\mu_{r_{_{2}}}}^{\left(s\right)}=\Omega_{\mu_{r_{_{1}}}\mu_{r_{_{2}}}}$.
In addition, $\left(r_{1},r_{2}\right)$ is an increasing sequence
constructed with the elements of $\left\{ 1,2,3,4\right\} \backslash\left\{ k_{1}^{s},k_{2}^{s}\right\} $}}.

As another example, the traceless-Ricci tensor $S_{\mu\nu}=\rho\lambda_{\mu}\lambda_{\nu}$
has the $\lambda$-weight 2.\textbf{\vspace{0.5cm}
}

\noindent \textbf{Definition 3 -- $\lambda$-conserving tensor:} Let
$E$ be a $\lambda$-weight $n$ tensor of rank $\left(0,m\right)$.
The $E$ tensor is $\lambda$-conserving if its $\lambda$-weight
increases by one after each nonzero contraction with $\lambda$.

\textbf{\vspace{0.2cm}
}

\noindent \textbf{Example 2:} $\nabla_{\mu_{1}}\xi_{\mu_{2}}$ is
a $\lambda$-weight conserving tensor since under a contraction with
one $\lambda$ vector, its $\lambda$-weight becomes 1 as
\begin{equation}
\lambda^{\mu_{1}}\nabla_{\mu_{1}}\xi_{\mu_{2}}=-\lambda_{\mu_{2}}\left(\frac{1}{4}\xi^{\mu_{1}}\xi_{\mu_{1}}-\frac{R}{D\left(D-1\right)}\right),
\end{equation}
and a further $\lambda$-contraction yields zero. Also, $\nabla_{\mu_{1}}\nabla_{\mu_{2}}\xi_{\mu_{3}}$
is $\lambda$-weight conserving since under a contraction with one
$\lambda$ vector, its $\lambda$-weight becomes 1 as can be seen
from all the nonvanishing contractions\textit{\emph{
\begin{align}
\lambda^{\mu_{1}}\nabla_{\mu_{1}}\nabla_{\mu_{2}}\xi_{\mu_{3}}= & -\frac{1}{2}\lambda_{\mu_{2}}\left[\xi_{\mu_{3}}\left(\frac{1}{4}\xi^{\mu_{1}}\xi_{\mu_{1}}-\frac{R}{D\left(D-1\right)}\right)+\xi^{\mu_{1}}\nabla_{\mu_{1}}\xi_{\mu_{3}}\right]\nonumber \\
 & +\lambda_{\mu_{3}}\left(-\frac{1}{2}\xi^{\mu_{1}}\nabla_{\mu_{2}}\xi_{\mu_{1}}+\frac{R}{D\left(D-1\right)}\xi_{\mu_{2}}\right),
\end{align}
\begin{equation}
\lambda^{\mu_{2}}\nabla_{\mu_{1}}\nabla_{\mu_{2}}\xi_{\mu_{3}}=-\frac{1}{2}\lambda_{\mu_{1}}\left[\xi_{\mu_{3}}\left(\frac{1}{4}\xi^{\mu_{2}}\xi_{\mu_{2}}-\frac{R}{D\left(D-1\right)}\right)+\xi^{\mu_{2}}\nabla_{\mu_{2}}\xi_{\mu_{3}}\right]-\frac{1}{2}\lambda_{\mu_{3}}\xi^{\mu_{2}}\nabla_{\mu_{1}}\xi_{\mu_{2}},
\end{equation}
\begin{equation}
\lambda^{\mu_{3}}\nabla_{\mu_{1}}\nabla_{\mu_{2}}\xi_{\mu_{3}}=-\frac{1}{2}\lambda_{\mu_{1}}\left(\xi^{\mu_{3}}\nabla_{\mu_{2}}\xi_{\mu_{3}}+\frac{1}{2}\xi_{\mu_{2}}\xi^{\mu_{3}}\xi_{\mu_{3}}\right)-\lambda_{\mu_{2}}\xi^{\mu_{3}}\nabla_{\mu_{1}}\xi_{\mu_{3}},
\end{equation}
}}and again a further $\lambda$-contraction with any of the above
contraction patterns of $\nabla_{\mu_{1}}\nabla_{\mu_{2}}\xi_{\mu_{3}}$
yields zero.\textbf{\vspace{0.5cm}
}

\noindent \textbf{Lemma 1: }\emph{For a $\lambda$-conserving tensor
of rank $\left(0,m\right)$ and $\lambda$-weight $n$, the maximum
number of nonzero $\lambda$ contractions is $\left(m-n\right)/2$
for even $\left(m-n\right)$ and $\left(m-n-1\right)/2$ for odd $\left(m-n\right)$.}\textbf{\vspace{0.2cm}
}

\noindent \textbf{Proof: }Under $p$ number of $\lambda$ contractions,
a $\lambda$-conserving $E$ tensor of $\lambda$-weight $n$ and
rank $\left(0,m\right)$ yields a $\lambda$-weight $\left(n+p\right)$
tensor of rank $\left(0,m-p\right)$ in the form
\begin{equation}
\lambda^{\mu_{j_{_{1}}}}\lambda^{\mu_{j_{_{2}}}}\dots\lambda^{\mu_{j_{_{p}}}}E_{\mu_{1}\mu_{2}\cdots\mu_{m}}=\sum_{s=1}^{N}\lambda_{\mu_{_{k_{_{1}}^{s}}}}\lambda_{\mu_{_{k_{_{2}}^{s}}}}\dots\lambda_{\mu_{_{k_{_{n+p}}^{s}}}}F_{\mu_{r_{_{1}}}\mu_{r_{_{2}}}\dots\mu_{r_{_{m-n-2p}}}}^{\left(s\right)},\label{eq:Multiple_l_cont_of_E}
\end{equation}
\textit{\emph{where }}$\left\{ j_{1},j_{2},\dots,j_{p}\right\} $
is a subset of $\left\{ 1,2,\dots,m\right\} $\textit{\emph{, $N$
is the number of the $\left(n+p\right)$-element subsets of $\left\{ 1,2,\dots,m\right\} \backslash\left\{ j_{1},j_{2},\dots,j_{p}\right\} $,
that is $N=\left(\begin{array}{c}
m-p\\
n+p
\end{array}\right)$, $s$ is the label for these $\left(n+p\right)$-element subsets
such that for each $s$, $\left\{ k_{1}^{s},k_{2}^{s},\dots,k_{n+p}^{s}\right\} $
is one of these subsets, and $\left(r_{1},r_{2},\dots,r_{m-n-2p}\right)$
is an increasing sequence constructed from $\left\{ 1,2,\dots,m\right\} \backslash\left(\left\{ j_{1},j_{2},\dots,j_{p}\right\} \cup\left\{ k_{1}^{s},k_{2}^{s},\dots,k_{n+p}^{s}\right\} \right)$.
Here, we assumed $m-n>2p$ as must be the case in (\ref{eq:Multiple_l_cont_of_E}).}}

\textit{\emph{This result implies that the maximum number of $\lambda$
contractions }}is $\left(m-n\right)/2$ for even $\left(m-n\right)$
and $\left(m-n-1\right)/2$ for odd $\left(m-n\right)$\textit{\emph{.
Then, one gets the following results, respectively, for even and odd
$\left(m-n\right)$;}}
\[
\left(\prod_{s=1}^{\frac{m-n}{2}}\lambda^{\mu_{j_{s}}}\right)E_{\mu_{1}\mu_{2}\cdots\mu_{m}}=\left(\prod_{s=1}^{\frac{m+n}{2}}\lambda_{\mu_{s}}\right)F,
\]
\textit{\emph{and}}
\[
\left(\prod_{s=1}^{\frac{m-n-1}{2}}\lambda^{\mu_{j_{s}}}\right)E_{\mu_{1}\mu_{2}\cdots\mu_{m}}=\sum_{s=1}^{\frac{m+n-1}{2}}F_{\mu_{s}}\lambda_{\mu_{r_{_{1}}}}\lambda_{\mu_{r_{_{2}}}}\dots\lambda_{\mu_{r_{_{\left(m+n-1\right)/2}}}},
\]
\textit{\emph{where $\left\{ r_{1},r_{2},\dots,r_{\left(m+n-1\right)/2}\right\} =\left\{ 1,2,\dots,m\right\} \backslash\left\{ j_{1},j_{2},\dots,j_{\left(m-n-1\right)/2},s\right\} $.
Here, note that for a $\lambda$-conserving tensor $E$, $\lambda^{\mu_{s}}F_{\mu_{s}}$
should be zero. This proves the lemma. \QEDA }}\textbf{\vspace{0.5cm}
}

\noindent \textbf{Lemma 2: }\emph{For a $\lambda$-conserving tensor
of rank $\left(0,m\right)$ and $\lambda$-weight $n$, the contractions
among its indices do not change the $\lambda$-weight of the tensor.}\textbf{\vspace{0.2cm}
}

\noindent \textbf{Proof: }A $\lambda$-weight $n$ tensor $E$ of
rank $\left(0,m\right)$ has the form
\[
E_{\mu_{1}\mu_{2}\cdots\mu_{m}}=\sum_{s=1}^{N}\lambda_{\mu_{_{k_{_{1}}^{s}}}}\lambda_{\mu_{_{k_{_{2}}^{s}}}}\dots\lambda_{\mu_{_{k_{_{n}}^{s}}}}F_{\mu_{r_{_{1}}}\mu_{r_{_{2}}}\dots\mu_{r_{_{m-n}}}}^{\left(s\right)},
\]
\textit{\emph{where $\left(r_{1},r_{2},\dots,r_{m-n}\right)$ is an
increasing sequence constructed with the elements of $\left\{ 1,2,\dots,m\right\} \backslash\left\{ k_{1}^{s},k_{2}^{s},\dots,k_{n}^{s}\right\} $.
The $\lambda$-weight zero tensors $F_{\mu_{r_{_{1}}}\mu_{r_{_{2}}}\dots\mu_{r_{_{m-n}}}}^{\left(s\right)}$
are $\lambda$-conserving since $E$ is $\lambda$-conserving. Then,
contractions among the indices of $E$ can be either $\lambda$-$\lambda$
contraction, or $\lambda$-$F$ contraction, or a contraction among
the indices of the $F$ tensor. The $\lambda$-$\lambda$ contraction
is zero. A contraction among the indices of the $F$ tensor surely
does not change the $\lambda$-weight. Finally, the result of each
$\lambda$-$F$ contraction increases the $\lambda$-weight by one,
so the total $\lambda$-weight still remains as $n$. \QEDA}}\textbf{\vspace{0.5cm}
}

\noindent \textbf{\textit{\emph{Theorem 2:}}} \textit{The rank $\left(0,n\right)$
tensor $\nabla^{n-1}\xi$ is $\lambda$-conserving.} \textbf{\vspace{0.2cm}
}

To prove this theorem, we need the following two lemmas below.\textit{\emph{
Let us introduce the indices of $\nabla^{n-1}\xi$ as
\begin{equation}
\left(\prod_{i=1}^{n-1}\nabla_{\mu_{i}}\right)\xi_{\mu_{n}}\equiv\nabla_{\mu_{1}}\nabla_{\mu_{2}}\dots\nabla_{\mu_{n-1}}\xi_{\mu_{n}}.
\end{equation}
To show that $\nabla^{n-1}\xi$ is $\lambda$-conserving, first let
us prove that $\lambda^{\mu_{j}}\left(\prod_{i=1}^{n-1}\nabla_{\mu_{i}}\right)\xi_{\mu_{n}}$,
where $j$ takes a value from $\left\{ 1,2,\dots,n\right\} $, is
$\lambda$-reducible by using mathematical induction.}}

\textbf{\vspace{0.5cm}
}

\noindent \textbf{Lemma 3:}\textbf{\emph{ }}\textit{$\lambda^{\mu_{j}}\left(\prod_{i=1}^{n-1}\nabla_{\mu_{i}}\right)\xi_{\mu_{n}}$,
where $j$ takes a value from $\left\{ 1,2,\dots,n\right\} $, is
$\lambda$-reducible.}\textbf{\vspace{0.2cm}
}

\noindent \textbf{Proof:}\textit{\emph{ As the basis set of identities,
we know that $\nabla_{\mu}\lambda_{\nu}=\xi_{(\mu}\lambda_{\nu)}$
and $\xi$ satisfies the identities }}
\begin{equation}
\lambda^{\mu_{1}}\nabla_{\mu_{1}}\xi_{\mu_{2}}=-\lambda_{\mu_{2}}\left(\frac{1}{4}\xi^{\mu_{1}}\xi_{\mu_{1}}-\frac{R}{D\left(D-1\right)}\right),
\end{equation}
and\textit{\emph{
\begin{equation}
\lambda^{\mu_{2}}\nabla_{\mu_{1}}\xi_{\mu_{2}}=-\frac{1}{2}\lambda_{\mu_{1}}\xi^{\mu_{2}}\xi_{\mu_{2}}.
\end{equation}
For mathematical induction, as the first step, the $n=2$ case given
above is sufficient. But, we will include the $n=3$ case, to obtain
some insight which will be useful in further calculations. Then, moving
to the mathematical induction proof:}}
\begin{enumerate}
\item \textit{\emph{For $n=3$, $\lambda^{\mu_{j}}\left(\prod_{i=1}^{2}\nabla_{\mu_{i}}\right)\xi_{\mu_{3}}$
involves the contraction patterns
\begin{equation}
\lambda^{\mu_{1}}\nabla_{\mu_{1}}\nabla_{\mu_{2}}\xi_{\mu_{3}},\qquad\lambda^{\mu_{2}}\nabla_{\mu_{1}}\nabla_{\mu_{2}}\xi_{\mu_{3}},\qquad\lambda^{\mu_{3}}\nabla_{\mu_{1}}\nabla_{\mu_{2}}\xi_{\mu_{3}}.
\end{equation}
The first contraction pattern reduces to the second one by interchanging
the order of the derivatives as
\begin{align}
\lambda^{\mu_{1}}\nabla_{\mu_{1}}\nabla_{\mu_{2}}\xi_{\mu_{3}} & =\lambda^{\mu_{1}}\left[\nabla_{\mu_{1}},\nabla_{\mu_{2}}\right]\xi_{\mu_{3}}+\lambda^{\mu_{1}}\nabla_{\mu_{2}}\nabla_{\mu_{1}}\xi_{\mu_{3}}\nonumber \\
 & =\lambda^{\mu_{1}}R_{\mu_{1}\mu_{2}\mu_{3}\mu_{4}}\xi^{\mu_{4}}+\lambda^{\mu_{1}}\nabla_{\mu_{2}}\nabla_{\mu_{1}}\xi_{\mu_{3}},
\end{align}
and from (B27) of \cite{gurses1}, one has
\begin{align}
\lambda^{\mu_{1}}\nabla_{\mu_{1}}\nabla_{\mu_{2}}\xi_{\mu_{3}} & =\frac{R}{D\left(D-1\right)}\lambda_{\mu_{3}}\xi_{\mu_{2}}+\lambda^{\mu_{1}}\nabla_{\mu_{2}}\nabla_{\mu_{1}}\xi_{\mu_{3}}.
\end{align}
Thus, if the second contraction pattern is $\lambda$-reducible, then
so is the first one. Moving to the second contraction pattern which
becomes
\begin{align}
\lambda^{\mu_{2}}\nabla_{\mu_{1}}\nabla_{\mu_{2}}\xi_{\mu_{3}}= & \nabla_{\mu_{1}}\left(\lambda^{\mu_{2}}\nabla_{\mu_{2}}\xi_{\mu_{3}}\right)-\left(\nabla_{\mu_{1}}\lambda^{\mu_{2}}\right)\nabla_{\mu_{2}}\xi_{\mu_{3}}\nonumber \\
= & \nabla_{\mu_{1}}\left(\lambda^{\mu_{2}}\nabla_{\mu_{2}}\xi_{\mu_{3}}\right)-\frac{1}{2}\xi_{\mu_{1}}\lambda^{\mu_{2}}\nabla_{\mu_{2}}\xi_{\mu_{3}}-\frac{1}{2}\lambda_{\mu_{1}}\xi^{\mu_{2}}\nabla_{\mu_{2}}\xi_{\mu_{3}},
\end{align}
and using the identity for $\lambda^{\mu_{2}}\nabla_{\mu_{2}}\xi_{\mu_{3}}$,
one obtains
\begin{equation}
\lambda^{\mu_{2}}\nabla_{\mu_{1}}\nabla_{\mu_{2}}\xi_{\mu_{3}}=-\frac{1}{2}\lambda_{\mu_{1}}\left[\xi_{\mu_{3}}\left(\frac{1}{4}\xi^{\mu_{2}}\xi_{\mu_{2}}-\frac{R}{D\left(D-1\right)}\right)+\xi^{\mu_{2}}\nabla_{\mu_{2}}\xi_{\mu_{3}}\right]-\frac{1}{2}\lambda_{\mu_{3}}\xi^{\mu_{2}}\nabla_{\mu_{1}}\xi_{\mu_{2}},\label{eq:O3_j2}
\end{equation}
which is $\lambda$-reducible. With this result, the first contraction
is also $\lambda$-reducible and takes the form
\begin{align}
\lambda^{\mu_{1}}\nabla_{\mu_{1}}\nabla_{\mu_{2}}\xi_{\mu_{3}}= & -\frac{1}{2}\lambda_{\mu_{2}}\left[\xi_{\mu_{3}}\left(\frac{1}{4}\xi^{\mu_{1}}\xi_{\mu_{1}}-\frac{R}{D\left(D-1\right)}\right)+\xi^{\mu_{1}}\nabla_{\mu_{1}}\xi_{\mu_{3}}\right]\nonumber \\
 & +\lambda_{\mu_{3}}\left(-\frac{1}{2}\xi^{\mu_{1}}\nabla_{\mu_{2}}\xi_{\mu_{1}}+\frac{R}{D\left(D-1\right)}\xi_{\mu_{2}}\right).\label{eq:O3_j1}
\end{align}
Lastly, the third contraction pattern can be written as
\begin{align}
\lambda^{\mu_{3}}\nabla_{\mu_{1}}\nabla_{\mu_{2}}\xi_{\mu_{3}}= & \nabla_{\mu_{1}}\left(\lambda^{\mu_{3}}\nabla_{\mu_{2}}\xi_{\mu_{3}}\right)-\left(\nabla_{\mu_{1}}\lambda^{\mu_{3}}\right)\nabla_{\mu_{2}}\xi_{\mu_{3}}\nonumber \\
= & \nabla_{\mu_{1}}\left(\lambda^{\mu_{3}}\nabla_{\mu_{2}}\xi_{\mu_{3}}\right)-\frac{1}{2}\xi_{\mu_{1}}\lambda^{\mu_{3}}\nabla_{\mu_{2}}\xi_{\mu_{3}}-\frac{1}{2}\lambda_{\mu_{1}}\xi^{\mu_{3}}\nabla_{\mu_{2}}\xi_{\mu_{3}},
\end{align}
and using the identity for $\lambda^{\mu_{2}}\nabla_{\mu_{1}}\xi_{\mu_{2}}$,
one obtains
\begin{equation}
\lambda^{\mu_{3}}\nabla_{\mu_{1}}\nabla_{\mu_{2}}\xi_{\mu_{3}}=-\frac{1}{2}\lambda_{\mu_{1}}\left(\xi^{\mu_{3}}\nabla_{\mu_{2}}\xi_{\mu_{3}}+\frac{1}{2}\xi_{\mu_{2}}\xi^{\mu_{3}}\xi_{\mu_{3}}\right)-\lambda_{\mu_{2}}\xi^{\mu_{3}}\nabla_{\mu_{1}}\xi_{\mu_{3}},\label{eq:O3_j3}
\end{equation}
which is also $\lambda$-reducible. In summary, $\lambda^{\mu_{j}}\left(\prod_{i=1}^{2}\nabla_{\mu_{i}}\right)\xi_{\mu_{3}}$
is $\lambda$-reducible as
\begin{equation}
\lambda^{\mu_{j}}\left(\prod_{i=1}^{2}\nabla_{\mu_{i}}\right)\xi_{\mu_{3}}=\sum_{\underset{\left(k\ne j\right)}{k=1}}^{3}\lambda_{\mu_{k}}E_{\mu_{m_{_{1}}}}^{\left(k,3,j\right)},\qquad m_{1}\in\left\{ 1,2,3\right\} \backslash\left\{ j,k\right\} ,
\end{equation}
where $E_{\mu_{m_{_{1}}}}^{\left(k,3,j\right)}$ are built from one-form
contractions of the building blocks;
\begin{equation}
\xi_{\mu_{1}},\qquad\nabla_{\mu_{1}}\xi_{\mu_{2}}.
\end{equation}
}}
\item \textit{\emph{Assume that $\lambda^{\mu_{j}}\left(\prod_{i=1}^{n-2}\nabla_{\mu_{i}}\right)\xi_{\mu_{n-1}}$
is $\lambda$-reducible for all $j$ in $\left\{ 1,2,\dots,n-1\right\} $
as
\begin{equation}
\lambda^{\mu_{j}}\left(\prod_{i=1}^{n-2}\nabla_{\mu_{i}}\right)\xi_{\mu_{n-1}}=\sum_{\underset{\left(k\ne j\right)}{k=1}}^{n-1}\lambda_{\mu_{k}}E_{\mu_{m_{_{1}}}\mu_{m_{_{2}}}\dots\mu_{m_{_{n-3}}}}^{\left(k,n-1,j\right)},\label{eq:Assumption}
\end{equation}
where $\left(m_{1},m_{2},\dots,m_{n-3}\right)$ is an }}\textit{increasing}\textit{\emph{
sequence constructed with the elements of $\left\{ 1,2,\dots,n-1\right\} \backslash\left\{ j,k\right\} $.
The $E_{\mu_{m_{_{1}}}\mu_{m_{_{2}}}\dots\mu_{m_{_{n-3}}}}^{\left(k,n-1,j\right)}$
tensors are built from the rank $\left(0,n-3\right)$ contractions
of the building blocks;
\begin{equation}
g_{\mu_{1}\mu_{2}},\qquad\xi_{\mu_{1}},\qquad\left(\prod_{i=1}^{r-1}\nabla_{\mu_{i}}\right)\xi_{\mu_{r}},\qquad r=2,3,\dots,n-2.
\end{equation}
}}
\item \textit{\emph{Then, we must show that (\ref{eq:Assumption}) holds
also for $n\rightarrow n+1$, that is $\lambda^{\mu_{j}}\left(\prod_{i=1}^{n-1}\nabla_{\mu_{i}}\right)\xi_{\mu_{n}}$
is $\lambda$-reducible for all $j$ in $\left\{ 1,2,\dots,n\right\} $
as
\begin{equation}
\lambda^{\mu_{j}}\left(\prod_{i=1}^{n-1}\nabla_{\mu_{i}}\right)\xi_{\mu_{n}}=\sum_{\underset{\left(k\ne j\right)}{k=1}}^{n}\lambda_{\mu_{k}}E_{\mu_{m_{_{1}}}\mu_{m_{_{2}}}\dots\mu_{m_{_{n-2}}}}^{\left(k,n,j\right)},\label{eq:One_l_cont_On}
\end{equation}
where $\left(m_{1},m_{2},\dots,m_{n-2}\right)$ is an }}\textit{increasing}\textit{\emph{
sequence constructed with the elements of $\left\{ 1,2,\dots,n\right\} \backslash\left\{ j,k\right\} $.
The $E_{\mu_{m_{_{1}}}\mu_{m_{_{2}}}\dots\mu_{m_{_{n-2}}}}^{\left(k,n,j\right)}$
tensors}}%
\footnote{\textit{\emph{It may seem that we label $E_{\mu_{m_{_{1}}}\mu_{m_{_{2}}}\dots\mu_{m_{_{n-2}}}}^{\left(k,n,j\right)}$
with the dummy index $\mu_{j}$, but in fact the $j$ label represents
the position of the covariant derivative whose index is contracted
with the index of the $\lambda$ vector. In this way, the $k$ label
represents the position of the index of $\lambda_{\mu_{_{k}}}$ between
the indices on the left-hand side.}}%
}\textit{\emph{ are built from the rank $\left(0,n-2\right)$ contractions
of the building blocks;
\begin{equation}
g_{\mu_{1}\mu_{2}},\qquad\xi_{\mu_{1}},\qquad\left(\prod_{i=1}^{r-1}\nabla_{\mu_{i}}\right)\xi_{\mu_{r}},\qquad r=2,3,\dots,n-1.
\end{equation}
To show this, first note that the contraction pattern for $j=1$,
that is
\begin{equation}
\lambda^{\mu_{1}}\nabla_{\mu_{1}}\left(\prod_{i=2}^{n-1}\nabla_{\mu_{i}}\right)\xi_{\mu_{n}},
\end{equation}
can be reduced to the $j=2$ term added with some terms involving
the $\left(n-2\right)$ order term $\nabla^{n-3}\xi$ after changing
the order of the first two covariant derivatives as
\begin{align}
\lambda^{\mu_{1}}\nabla_{\mu_{1}}\nabla_{\mu_{2}}\left(\prod_{i=3}^{n-1}\nabla_{\mu_{i}}\right)\xi_{\mu_{n}}= & \lambda^{\mu_{1}}\left[\nabla_{\mu_{1}},\nabla_{\mu_{2}}\right]\left(\prod_{i=3}^{n-1}\nabla_{\mu_{i}}\right)\xi_{\mu_{n}}+\lambda^{\mu_{1}}\nabla_{\mu_{2}}\nabla_{\mu_{1}}\left(\prod_{i=3}^{n-1}\nabla_{\mu_{i}}\right)\xi_{\mu_{n}}\nonumber \\
= & \lambda^{\mu_{1}}\sum_{s=3}^{n-1}R_{\mu_{1}\mu_{2}\mu_{s}}^{\phantom{\mu_{1}\mu_{2}\mu_{s}}\mu_{n+1}}\left(\prod_{i_{1}=3}^{s-1}\nabla_{\mu_{i_{_{1}}}}\right)\nabla_{\mu_{n+1}}\left(\prod_{i_{2}=s+1}^{n-1}\nabla_{\mu_{i_{_{2}}}}\right)\xi_{\mu_{n}}\nonumber \\
 & +\lambda^{\mu_{1}}R_{\mu_{1}\mu_{2}\mu_{n}}^{\phantom{\mu_{1}\mu_{2}\mu_{s}}\mu_{n+1}}\left(\prod_{i=3}^{n-1}\nabla_{\mu_{i}}\right)\xi_{\mu_{n+1}}\nonumber \\
 & +\lambda^{\mu_{1}}\nabla_{\mu_{2}}\nabla_{\mu_{1}}\left(\prod_{i=3}^{n-1}\nabla_{\mu_{i}}\right)\xi_{\mu_{n}},
\end{align}
where using $\lambda^{\mu_{1}}R_{\mu_{1}\mu_{2}\mu_{3}\mu_{4}}=\frac{R}{D\left(D-1\right)}\left(\lambda_{\mu_{3}}g_{\mu_{2}\mu_{4}}-\lambda_{\mu_{4}}g_{\mu_{2}\mu_{3}}\right)$,
one arrives at
\begin{align}
 & \lambda^{\mu_{1}}\nabla_{\mu_{1}}\nabla_{\mu_{2}}\left(\prod_{i=3}^{n-1}\nabla_{\mu_{i}}\right)\xi_{\mu_{n}}\nonumber \\
= & \frac{R}{D\left(D-1\right)}\left[\lambda_{\mu_{n}}\left(\prod_{i=3}^{n-1}\nabla_{\mu_{i}}\right)\xi_{\mu_{2}}+\sum_{s=3}^{n-1}\lambda_{\mu_{s}}\left(\prod_{i_{1}=3}^{s-1}\nabla_{\mu_{i_{_{1}}}}\right)\nabla_{\mu_{2}}\left(\prod_{i_{2}=s+1}^{n-1}\nabla_{\mu_{i_{_{2}}}}\right)\xi_{\mu_{n}}\right]\nonumber \\
 & -\frac{R}{D\left(D-1\right)}g_{\mu_{2}\mu_{n}}\lambda^{\mu_{n+1}}\left(\prod_{i=3}^{n-1}\nabla_{\mu_{i}}\right)\xi_{\mu_{n+1}}\nonumber \\
 & -\frac{R}{D\left(D-1\right)}\sum_{s=3}^{n-1}g_{\mu_{2}\mu_{s}}\lambda^{\mu_{n+1}}\left(\prod_{i_{1}=3}^{s-1}\nabla_{\mu_{i_{_{1}}}}\right)\nabla_{\mu_{n+1}}\left(\prod_{i_{2}=s+1}^{n-1}\nabla_{\mu_{i_{_{2}}}}\right)\xi_{\mu_{n}}\nonumber \\
 & +\lambda^{\mu_{1}}\nabla_{\mu_{2}}\nabla_{\mu_{1}}\left(\prod_{i=3}^{n-1}\nabla_{\mu_{i}}\right)\xi_{\mu_{n}}.\label{eq:1st_cont_pat_On}
\end{align}
The first line is $\lambda$-reducible and involves the $n-2$ order
term $\nabla^{n-3}\xi$, the second and the third lines involve all
one-$\lambda$ contraction patterns of $\nabla^{n-3}\xi$, that is
\begin{equation}
\lambda^{\nu_{j}}\left(\prod_{i=1}^{n-3}\nabla_{\nu_{i}}\right)\xi_{\nu_{n-2}},
\end{equation}
with all possible $j$'s from $1$ to $n-2$, and the last line is
the $j=2$ contraction pattern of $\nabla^{n-1}\xi$. Assuming $\lambda^{\nu_{j}}\left(\prod_{i=1}^{n-3}\nabla_{\nu_{i}}\right)\xi_{\nu_{n-2}}$
is also $\lambda$-reducible, then the $j=1$ contraction pattern
of $\nabla^{n-1}\xi$ is $\lambda$-reducible if and only if the $j=2$
contraction pattern of $\nabla^{n-1}\xi$ is $\lambda$-reducible.
Let us move on to the analysis of the $1<j\le n$ contraction patterns
of $\nabla^{n-1}\xi$ and let us write
\begin{align}
\lambda^{\mu_{j}}\nabla_{\mu_{1}}\left(\prod_{i=2}^{n-1}\nabla_{\mu_{i}}\right)\xi_{\mu_{n}}= & \nabla_{\mu_{1}}\left[\lambda^{\mu_{j}}\left(\prod_{i=2}^{n-1}\nabla_{\mu_{i}}\right)\xi_{\mu_{n}}\right]-\left(\nabla_{\mu_{1}}\lambda^{\mu_{j}}\right)\left(\prod_{i=2}^{n-1}\nabla_{\mu_{i}}\right)\xi_{\mu_{n}}\nonumber \\
= & \nabla_{\mu_{1}}\left[\lambda^{\mu_{j}}\left(\prod_{i=2}^{n-1}\nabla_{\mu_{i}}\right)\xi_{\mu_{n}}\right]\nonumber \\
 & -\frac{1}{2}\xi_{\mu_{1}}\lambda^{\mu_{j}}\left(\prod_{i=2}^{n-1}\nabla_{\mu_{i}}\right)\xi_{\mu_{n}}-\frac{1}{2}\lambda_{\mu_{1}}\xi^{\mu_{j}}\left(\prod_{i=2}^{n-1}\nabla_{\mu_{i}}\right)\xi_{\mu_{n}},
\end{align}
where the last term is already $\lambda$-reducible while the first
and the second terms involve the order $n-1$ term $\nabla^{n-2}\xi$
which, from the assumption (\ref{eq:Assumption}), has the form}}%
\footnote{\textit{\emph{Note that in this form, we only updated the superscript
of $E^{\left(k,n-1,j\right)}$ to $E^{\left(k-1,n-1,j-1\right)}$
during the change in the range of $i$ (and so in $k$), because the
first and the third labels of $E^{\left(k,n-1,j\right)}$ correspond
to the position of the contracted index and the position of the index
of $\lambda_{\mu_{_{k}}}$ between the indices on the left-hand side.
With this update, the labeling still corresponds to the correct terms
in the lower order term. This enabled us to relate $E^{\left(k,n,j\right)}$'s
to $E^{\left(k,n-1,j\right)}$'s.}}%
}\textit{\emph{
\begin{equation}
\lambda^{\mu_{j}}\left(\prod_{i=2}^{n-1}\nabla_{\mu_{i}}\right)\xi_{\mu_{n}}=\sum_{\underset{\left(k\ne j\right)}{k=2}}^{n}\lambda_{\mu_{k}}E_{\mu_{m_{_{1}}}\mu_{m_{_{2}}}\dots\mu_{m_{_{n-3}}}}^{\left(k-1,n-1,j-1\right)},
\end{equation}
where $\left(m_{1},m_{2},\dots,m_{n-3}\right)$ is an increasing sequence
constructed from $\left\{ 2,3,\dots,n\right\} \backslash\left\{ j,k\right\} $.
Using this form, one finds
\begin{align}
\lambda^{\mu_{j}}\nabla_{\mu_{1}}\left(\prod_{i=2}^{n-1}\nabla_{\mu_{i}}\right)\xi_{\mu_{n}}= & \frac{1}{2}\lambda_{\mu_{1}}\left[\sum_{\underset{\left(k\ne j\right)}{k=2}}^{n}\xi_{\mu_{k}}E_{\mu_{m_{_{1}}}\mu_{m_{_{2}}}\dots\mu_{m_{_{n-3}}}}^{\left(k-1,n-1,j-1\right)}-\xi^{\mu_{j}}\left(\prod_{i=2}^{n-1}\nabla_{\mu_{i}}\right)\xi_{\mu_{n}}\right]\nonumber \\
 & +\sum_{\underset{\left(k\ne j\right)}{k=2}}^{n}\lambda_{\mu_{k}}\nabla_{\mu_{1}}E_{\mu_{m_{_{1}}}\mu_{m_{_{2}}}\dots\mu_{m_{_{n-3}}}}^{\left(k-1,n-1,j-1\right)},\label{eq:j_in_2_to_n_pattern}
\end{align}
so the $1<j\le n$ contraction patterns of $\nabla^{n-1}\xi$ are
$\lambda$-reducible. In addition, the $\lambda$-reducibility of
the $j=2$ pattern implies the $\lambda$-reducibility of the $j=1$
pattern.\QEDA}}
\end{enumerate}
\textbf{\vspace{0.5cm}
}

\noindent \textbf{\textit{\emph{Lemma 4:}}}\textbf{\textit{ }}\textit{The
$E_{\mu_{r_{_{1}}}\mu_{r_{_{2}}}\dots\mu_{r_{_{n-2}}}}^{\left(k,n,j\right)}$
tensors can be recursively obtained from the $E$ tensors of the lower
orders.}

\textbf{\vspace{0.2cm}
}

\noindent \textbf{\textit{\emph{Proof: }}}\textit{\emph{For the $1<j\le n$
contraction patterns of $\nabla^{n-1}\xi$, we just need to compare
(\ref{eq:j_in_2_to_n_pattern}) with
\begin{equation}
\lambda^{\mu_{j}}\left(\prod_{i=1}^{n-1}\nabla_{\mu_{i}}\right)\xi_{\mu_{n}}=\sum_{\underset{\left(k\ne j\right)}{k=1}}^{n}\lambda_{\mu_{k}}E_{\mu_{r_{_{1}}}\mu_{r_{_{2}}}\dots\mu_{r_{_{n-2}}}}^{\left(k,n,j\right)},
\end{equation}
where $\left(r_{1},r_{2},\dots,r_{n-2}\right)$ is an increasing sequence
constructed from $\left\{ 1,2,\dots,n\right\} \backslash\left\{ j,k\right\} $,
one finds
\begin{equation}
E_{\mu_{r_{_{1}}}\mu_{r_{_{2}}}\dots\mu_{r_{_{n-2}}}}^{\left(1,n,j\right)}=\frac{1}{2}\sum_{\underset{\left(k\ne j\right)}{k=2}}^{n}\xi_{\mu_{k}}E_{\mu_{m_{_{1}}}\mu_{m_{_{2}}}\dots\mu_{m_{_{n-3}}}}^{\left(k-1,n-1,j-1\right)}-\frac{1}{2}\xi^{\mu_{j}}\left(\prod_{i=2}^{n-1}\nabla_{\mu_{i}}\right)\xi_{\mu_{n}},\label{eq:Rec_rel_j_k1}
\end{equation}
\begin{equation}
k\ge2\quad\Rightarrow\quad E_{\mu_{r_{_{1}}}\mu_{r_{_{2}}}\dots\mu_{r_{_{n-2}}}}^{\left(k,n,j\right)}=\nabla_{\mu_{1}}E_{\mu_{m_{_{1}}}\mu_{m_{_{2}}}\dots\mu_{m_{_{n-3}}}}^{\left(k-1,n-1,j-1\right)},\label{eq:Rec_rel_j_k}
\end{equation}
where $\left(m_{1},m_{2},\dots,m_{n-3}\right)$ is an increasing sequence
constructed from $\left\{ 2,3,\dots,n\right\} \backslash\left\{ j,k\right\} $. }}

\textit{\emph{For the $j=1$ contraction pattern of $\nabla^{n-1}\xi$,}}\textit{
}\textit{\emph{we need to make the $\lambda$-reducibility assumption
for the $\left(n-2\right)^{{\rm th}}$ order term $\nabla^{n-3}\xi$
more explicit and assume the form
\begin{equation}
\lambda^{\mu_{j}}\left(\prod_{i=1}^{n-3}\nabla_{\mu_{i}}\right)\xi_{\mu_{n-2}}=\sum_{\underset{\left(k\ne j\right)}{k=1}}^{n-2}\lambda_{\mu_{k}}E_{\mu_{m_{_{1}}}\mu_{m_{_{2}}}\dots\mu_{m_{_{n-4}}}}^{\left(k,n-2,j\right)},\label{eq:On-2}
\end{equation}
where $\left(m_{1},m_{2},\dots,m_{n-4}\right)$ is an increasing sequence
constructed from $\left\{ 1,2,\dots,n-2\right\} \backslash\left\{ j,k\right\} $.
The rank $\left(0,n-4\right)$ tensors $E_{\mu_{m_{_{1}}}\mu_{m_{_{2}}}\dots\mu_{m_{_{n-4}}}}^{\left(k,n-2,j\right)}$
are built from the rank $\left(0,n-4\right)$ contractions of the
building blocks;
\begin{equation}
g_{\mu_{1}\mu_{2}},\qquad\xi_{\mu_{1}},\qquad\left(\prod_{i=1}^{r-1}\nabla_{\mu_{i}}\right)\xi_{\mu_{r}},\qquad r=2,3,\dots,n-3.
\end{equation}
Then, the terms
\begin{equation}
\lambda^{\mu_{n+1}}\left(\prod_{i_{1}=3}^{s-1}\nabla_{\mu_{i_{_{1}}}}\right)\nabla_{\mu_{n+1}}\left(\prod_{i_{2}=s+1}^{n-1}\nabla_{\mu_{i_{_{2}}}}\right)\xi_{\mu_{n}}\Rightarrow\lambda^{\mu_{s}}\left(\prod_{i=3}^{n-1}\nabla_{\mu_{i}}\right)\xi_{\mu_{n}},
\end{equation}
{[}where we used ``$\Rightarrow$'' with the meaning ``can be considered
as'' because it is not possible to put the right-hand-side term back
into (\ref{eq:1st_cont_pat_On}); however, considering it makes sense
as we just want to use (\ref{eq:On-2}),{]} and
\begin{equation}
\lambda^{\mu_{n+1}}\left(\prod_{i=3}^{n-1}\nabla_{\mu_{i}}\right)\xi_{\mu_{n+1}}=\lambda^{\mu_{s}}\left(\prod_{i=3}^{n-1}\nabla_{\mu_{i}}\right)\xi_{\mu_{s}},
\end{equation}
appearing in the $j=1$ contraction pattern of $\nabla^{n-1}\xi$,
that is (\ref{eq:1st_cont_pat_On}), can be written as
\begin{equation}
\lambda^{\mu_{s}}\left(\prod_{i=3}^{n-1}\nabla_{\mu_{i}}\right)\xi_{\mu_{n}}=\sum_{\underset{\left(k\ne s\right)}{k=3}}^{n}\lambda_{\mu_{k}}E_{\mu_{t_{_{1}}}\mu_{t_{_{2}}}\dots\mu_{t_{_{n-4}}}}^{\left(k-2,n-2,s-2\right)},\label{eq:On-2_shifted_range}
\end{equation}
where $\left(t_{1},t_{2},\dots,t_{n-4}\right)$ is an increasing sequence
constructed from $\left\{ 3,4,\dots,n\right\} \backslash\left\{ s,k\right\} $.
In addition, using the result for the $j=2$ contraction pattern of
$\nabla^{n-1}\xi$, that is
\begin{align}
\lambda^{\mu_{2}}\nabla_{\mu_{1}}\nabla_{\mu_{2}}\left(\prod_{i=3}^{n-1}\nabla_{\mu_{i}}\right)\xi_{\mu_{n}}= & \frac{1}{2}\lambda_{\mu_{1}}\left[\sum_{k=3}^{n}\xi_{\mu_{k}}E_{\mu_{m_{_{1}}}\mu_{m_{_{2}}}\dots\mu_{m_{_{n-3}}}}^{\left(k-1,n-1,1\right)}-\xi^{\mu_{2}}\nabla_{\mu_{2}}\left(\prod_{i=3}^{n-1}\nabla_{\mu_{i}}\right)\xi_{\mu_{n}}\right]\nonumber \\
 & +\sum_{k=3}^{n}\lambda_{\mu_{k}}\nabla_{\mu_{1}}E_{\mu_{m_{_{1}}}\mu_{m_{_{2}}}\dots\mu_{m_{_{n-3}}}}^{\left(k-1,n-1,1\right)},
\end{align}
where $\left(m_{1},m_{2},\dots,m_{n-3}\right)$ is an increasing sequence
constructed from $\left\{ 3,4,\dots,n\right\} \backslash\left\{ k\right\} $,
the last term in (\ref{eq:1st_cont_pat_On}) can be written as
\begin{align}
\lambda^{\mu_{1}}\nabla_{\mu_{2}}\nabla_{\mu_{1}}\left(\prod_{i=3}^{n-1}\nabla_{\mu_{i}}\right)\xi_{\mu_{n}}= & \frac{1}{2}\lambda_{\mu_{2}}\left[\sum_{k=3}^{n}\xi_{\mu_{k}}E_{\mu_{m_{_{1}}}\mu_{m_{_{2}}}\dots\mu_{m_{_{n-3}}}}^{\left(k-1,n-1,1\right)}-\xi^{\mu_{1}}\nabla_{\mu_{1}}\left(\prod_{i=3}^{n-1}\nabla_{\mu_{i}}\right)\xi_{\mu_{n}}\right]\nonumber \\
 & +\sum_{k=3}^{n}\lambda_{\mu_{k}}\nabla_{\mu_{2}}E_{\mu_{m_{_{1}}}\mu_{m_{_{2}}}\dots\mu_{m_{_{n-3}}}}^{\left(k-1,n-1,1\right)},\label{eq:2nd_cont_pat_On}
\end{align}
where $\left(m_{1},m_{2},\dots,m_{n-3}\right)$ is an increasing sequence
constructed from $\left\{ 3,4,\dots,n\right\} \backslash\left\{ k\right\} $.
Using (\ref{eq:On-2_shifted_range}) and (\ref{eq:2nd_cont_pat_On})
in (\ref{eq:1st_cont_pat_On}), one obtains
\begin{align}
 & \lambda^{\mu_{1}}\nabla_{\mu_{1}}\nabla_{\mu_{2}}\left(\prod_{i=3}^{n-1}\nabla_{\mu_{i}}\right)\xi_{\mu_{n}}\nonumber \\
= & \frac{R}{D\left(D-1\right)}\left[\lambda_{\mu_{n}}\left(\prod_{i=3}^{n-1}\nabla_{\mu_{i}}\right)\xi_{\mu_{2}}+\sum_{s=3}^{n-1}\lambda_{\mu_{s}}\left(\prod_{i_{1}=3}^{s-1}\nabla_{\mu_{i_{_{1}}}}\right)\nabla_{\mu_{2}}\left(\prod_{i_{2}=s+1}^{n-1}\nabla_{\mu_{i_{_{2}}}}\right)\xi_{\mu_{n}}\right]\nonumber \\
 & -\frac{R}{D\left(D-1\right)}\left[g_{\mu_{2}\mu_{n}}\sum_{k=3}^{n-1}\lambda_{\mu_{k}}E_{\mu_{t_{_{1}}}\mu_{t_{_{2}}}\dots\mu_{t_{_{n-4}}}}^{\left(k-2,n-2,n-2\right)}+\sum_{s=3}^{n-1}g_{\mu_{2}\mu_{s}}\sum_{\underset{\left(k\ne s\right)}{k=3}}^{n}\lambda_{\mu_{k}}E_{\mu_{t_{_{1}}}\mu_{t_{_{2}}}\dots\mu_{t_{_{n-4}}}}^{\left(k-2,n-2,s-2\right)}\right]\nonumber \\
 & +\frac{1}{2}\lambda_{\mu_{2}}\left[\sum_{k=3}^{n}\xi_{\mu_{k}}E_{\mu_{m_{_{1}}}\mu_{m_{_{2}}}\dots\mu_{m_{_{n-3}}}}^{\left(k-1,n-1,1\right)}-\xi^{\mu_{1}}\nabla_{\mu_{1}}\left(\prod_{i=3}^{n-1}\nabla_{\mu_{i}}\right)\xi_{\mu_{n}}\right]\nonumber \\
 & +\sum_{k=3}^{n}\lambda_{\mu_{k}}\nabla_{\mu_{2}}E_{\mu_{m_{_{1}}}\mu_{m_{_{2}}}\dots\mu_{m_{_{n-3}}}}^{\left(k-1,n-1,1\right)}.
\end{align}
Here, we can change the order of summations in the second term of
the second line as
\begin{align}
\sum_{s=3}^{n-1}g_{\mu_{2}\mu_{s}}\sum_{\underset{\left(k\ne s\right)}{k=3}}^{n}\lambda_{\mu_{k}}E_{\mu_{t_{_{1}}}\mu_{t_{_{2}}}\dots\mu_{t_{_{n-4}}}}^{\left(k-2,n-2,s-2\right)}= & \sum_{k=3}^{n}\lambda_{\mu_{k}}\sum_{\underset{s\ne k}{s=3}}^{n-1}g_{\mu_{2}\mu_{s}}E_{\mu_{t_{_{1}}}\mu_{t_{_{2}}}\dots\mu_{t_{_{n-4}}}}^{\left(k-2,n-2,s-2\right)},
\end{align}
then with this result, one has
\begin{align}
 & \lambda^{\mu_{1}}\nabla_{\mu_{1}}\nabla_{\mu_{2}}\left(\prod_{i=3}^{n-1}\nabla_{\mu_{i}}\right)\xi_{\mu_{n}}\nonumber \\
= & \frac{1}{2}\lambda_{\mu_{2}}\left[\sum_{k=3}^{n}\xi_{\mu_{k}}E_{\mu_{m_{_{1}}}\mu_{m_{_{2}}}\dots\mu_{m_{_{n-3}}}}^{\left(k-1,n-1,1\right)}-\xi^{\mu_{1}}\nabla_{\mu_{1}}\left(\prod_{i=3}^{n-1}\nabla_{\mu_{i}}\right)\xi_{\mu_{n}}\right]\nonumber \\
 & +\sum_{k=3}^{n-1}\lambda_{\mu_{k}}\left[\frac{R}{D\left(D-1\right)}\left(\prod_{i_{1}=3}^{k-1}\nabla_{\mu_{i_{_{1}}}}\right)\nabla_{\mu_{2}}\left(\prod_{i_{2}=k+1}^{n-1}\nabla_{\mu_{i_{_{2}}}}\right)\xi_{\mu_{n}}+\nabla_{\mu_{2}}E_{\mu_{m_{_{1}}}\mu_{m_{_{2}}}\dots\mu_{m_{_{n-3}}}}^{\left(k-1,n-1,1\right)}\right]\nonumber \\
 & -\frac{R}{D\left(D-1\right)}\sum_{k=3}^{n-1}\lambda_{\mu_{k}}\left(g_{\mu_{2}\mu_{n}}E_{\mu_{t_{_{1}}}\mu_{t_{_{2}}}\dots\mu_{t_{_{n-4}}}}^{\left(k-2,n-2,n-2\right)}+\sum_{\underset{\left(s\ne k\right)}{s=3}}^{n-1}g_{\mu_{2}\mu_{s}}E_{\mu_{t_{_{1}}}\mu_{t_{_{2}}}\dots\mu_{t_{_{n-4}}}}^{\left(k-2,n-2,s-2\right)}\right)\nonumber \\
 & +\lambda_{\mu_{n}}\left\{ \frac{R}{D\left(D-1\right)}\left[\left(\prod_{i=3}^{n-1}\nabla_{\mu_{i}}\right)\xi_{\mu_{2}}-\sum_{s=3}^{n-1}g_{\mu_{2}\mu_{s}}E_{\mu_{t_{_{1}}}\mu_{t_{_{2}}}\dots\mu_{t_{_{n-4}}}}^{\left(n-2,n-2,s-2\right)}\right]+\nabla_{\mu_{2}}E_{\mu_{m_{_{1}}}\mu_{m_{_{2}}}\dots\mu_{m_{_{n-3}}}}^{\left(n-1,n-1,1\right)}\right\} ,
\end{align}
The third line of this result can be written as
\begin{equation}
g_{\mu_{2}\mu_{n}}E_{\mu_{t_{_{1}}}\mu_{t_{_{2}}}\dots\mu_{t_{_{n-4}}}}^{\left(k-2,n-2,n-2\right)}+\sum_{\underset{\left(s\ne k\right)}{s=3}}^{n-1}g_{\mu_{2}\mu_{s}}E_{\mu_{t_{_{1}}}\mu_{t_{_{2}}}\dots\mu_{t_{_{n-4}}}}^{\left(k-2,n-2,s-2\right)}=\sum_{\underset{\left(s\ne k\right)}{s=3}}^{n}g_{\mu_{2}\mu_{s}}E_{\mu_{t_{_{1}}}\mu_{t_{_{2}}}\dots\mu_{t_{_{n-4}}}}^{\left(k-2,n-2,s-2\right)},
\end{equation}
and then reordering the terms and rewriting the $t$ indices as $m$
indices, the final form becomes
\begin{align}
 & \lambda^{\mu_{1}}\nabla_{\mu_{1}}\nabla_{\mu_{2}}\left(\prod_{i=3}^{n-1}\nabla_{\mu_{i}}\right)\xi_{\mu_{n}}\nonumber \\
= & \frac{1}{2}\lambda_{\mu_{2}}\left[\sum_{k=3}^{n}\xi_{\mu_{k}}E_{\mu_{m_{_{1}}}\mu_{m_{_{2}}}\dots\mu_{m_{_{n-3}}}}^{\left(k-1,n-1,1\right)}-\xi^{\mu_{1}}\nabla_{\mu_{1}}\left(\prod_{i=3}^{n-1}\nabla_{\mu_{i}}\right)\xi_{\mu_{n}}\right]\nonumber \\
 & +\frac{R}{D\left(D-1\right)}\sum_{k=3}^{n-1}\lambda_{\mu_{k}}\left[\left(\prod_{i_{1}=3}^{k-1}\nabla_{\mu_{i_{_{1}}}}\right)\nabla_{\mu_{2}}\left(\prod_{i_{2}=k+1}^{n-1}\nabla_{\mu_{i_{_{2}}}}\right)\xi_{\mu_{n}}-\sum_{\underset{\left(s\ne k\right)}{s=3}}^{n}g_{\mu_{2}\mu_{s}}E_{\mu_{m_{_{1}}}\mu_{m_{_{2}}}\dots\mu_{m_{_{n-4}}}}^{\left(k-2,n-2,s-2\right)}\right]\nonumber \\
 & +\sum_{k=3}^{n-1}\lambda_{\mu_{k}}\nabla_{\mu_{2}}E_{\mu_{m_{_{1}}}\mu_{m_{_{2}}}\dots\mu_{m_{_{n-3}}}}^{\left(k-1,n-1,1\right)}\nonumber \\
 & +\lambda_{\mu_{n}}\left\{ \frac{R}{D\left(D-1\right)}\left[\left(\prod_{i=3}^{n-1}\nabla_{\mu_{i}}\right)\xi_{\mu_{2}}-\sum_{s=3}^{n-1}g_{\mu_{2}\mu_{s}}E_{\mu_{m_{_{1}}}\mu_{m_{_{2}}}\dots\mu_{m_{_{n-4}}}}^{\left(n-2,n-2,s-2\right)}\right]+\nabla_{\mu_{2}}E_{\mu_{m_{_{1}}}\mu_{m_{_{2}}}\dots\mu_{m_{_{n-3}}}}^{\left(n-1,n-1,1\right)}\right\} ,
\end{align}
where $\left(m_{1},m_{2},\dots,m_{n-3}\right)$ and $\left(m_{1},m_{2},\dots,m_{n-4}\right)$
are increasing sequences constructed from$\left\{ 3,4,\dots,n\right\} \backslash\left\{ k\right\} $
and $\left\{ 3,4,\dots,n\right\} \backslash\left\{ s,k\right\} $,
respectively.}}%
\footnote{\textit{\emph{Note that in the last line $k=n$.}}%
}\textit{\emph{ Then, by comparing this result with
\begin{equation}
\lambda^{\mu_{1}}\left(\prod_{i=1}^{n-1}\nabla_{\mu_{i}}\right)\xi_{\mu_{n}}=\sum_{k=2}^{n}\lambda_{\mu_{k}}E_{\mu_{r_{_{1}}}\mu_{r_{_{2}}}\dots\mu_{r_{_{n-2}}}}^{\left(k,n,1\right)},
\end{equation}
where $\left(r_{1},r_{2},\dots,r_{n-2}\right)$ is an increasing sequence
constructed from $\left\{ 2,3,\dots,n\right\} \backslash\left\{ k\right\} $,
one arrives at
\begin{equation}
E_{\mu_{r_{_{1}}}\mu_{r_{_{2}}}\dots\mu_{r_{_{n-2}}}}^{\left(2,n,1\right)}=\frac{1}{2}\left[\sum_{k=3}^{n}\xi_{\mu_{k}}E_{\mu_{m_{_{1}}}\mu_{m_{_{2}}}\dots\mu_{m_{_{n-3}}}}^{\left(k-1,n-1,1\right)}-\xi^{\mu_{1}}\nabla_{\mu_{1}}\left(\prod_{i=3}^{n-1}\nabla_{\mu_{i}}\right)\xi_{\mu_{n}}\right],\label{eq:Rec_rel_j1_k2}
\end{equation}
\begin{align}
2<k<n\quad\Rightarrow\quad E_{\mu_{r_{_{1}}}\mu_{r_{_{2}}}\dots\mu_{r_{_{n-2}}}}^{\left(k,n,1\right)}= & \frac{R}{D\left(D-1\right)}\left(\prod_{i_{1}=3}^{k-1}\nabla_{\mu_{i_{_{1}}}}\right)\nabla_{\mu_{2}}\left(\prod_{i_{2}=k+1}^{n-1}\nabla_{\mu_{i_{_{2}}}}\right)\xi_{\mu_{n}}\nonumber \\
 & -\frac{R}{D\left(D-1\right)}\sum_{\underset{\left(s\ne k\right)}{s=3}}^{n}g_{\mu_{2}\mu_{s}}E_{\mu_{m_{_{1}}}\mu_{m_{_{2}}}\dots\mu_{m_{_{n-4}}}}^{\left(k-2,n-2,s-2\right)}\nonumber \\
 & +\nabla_{\mu_{2}}E_{\mu_{m_{_{1}}}\mu_{m_{_{2}}}\dots\mu_{m_{_{n-3}}}}^{\left(k-1,n-1,1\right)},\label{eq:Rec_rel_j1_k}
\end{align}
\begin{align}
E_{\mu_{r_{_{1}}}\mu_{r_{_{2}}}\dots\mu_{r_{_{n-2}}}}^{\left(n,n,1\right)}= & \frac{R}{D\left(D-1\right)}\left[\left(\prod_{i=3}^{n-1}\nabla_{\mu_{i}}\right)\xi_{\mu_{2}}-\sum_{s=3}^{n-1}g_{\mu_{2}\mu_{s}}E_{\mu_{m_{_{1}}}\mu_{m_{_{2}}}\dots\mu_{m_{_{n-4}}}}^{\left(n-2,n-2,s-2\right)}\right]\nonumber \\
 & +\nabla_{\mu_{2}}E_{\mu_{m_{_{1}}}\mu_{m_{_{2}}}\dots\mu_{m_{_{n-3}}}}^{\left(n-1,n-1,1\right)}.\label{eq:Rec_rel_j1_kn}
\end{align}
where $\left(m_{1},m_{2},\dots,m_{n-3}\right)$ and $\left(m_{1},m_{2},\dots,m_{n-4}\right)$
are increasing sequences constructed from $\left\{ 3,4,\dots,n\right\} \backslash\left\{ k\right\} $
and $\left\{ 3,4,\dots,n\right\} \backslash\left\{ s,k\right\} $,
respectively. \QEDA}}

\textbf{\vspace{0.2cm}
}

\noindent \textbf{\textit{\emph{Example 3: }}}\textit{\emph{To apply
the recursive relations (\ref{eq:Rec_rel_j_k1}) and (\ref{eq:Rec_rel_j_k}),
let us use the $n=3$ result. For $j=2$, one has
\begin{align}
\lambda^{\mu_{2}}\left(\prod_{i=1}^{2}\nabla_{\mu_{i}}\right)\xi_{\mu_{3}} & =\sum_{\underset{\left(k\ne2\right)}{k=1}}^{3}\lambda_{\mu_{k}}E_{\mu_{r_{_{1}}}}^{\left(k,3,2\right)}\nonumber \\
\lambda^{\mu_{2}}\nabla_{\mu_{1}}\nabla_{\mu_{2}}\xi_{\mu_{3}} & =\lambda_{\mu_{1}}E_{\mu_{3}}^{\left(1,3,2\right)}+\lambda_{\mu_{3}}E_{\mu_{1}}^{\left(3,3,2\right)},
\end{align}
and from (\ref{eq:Rec_rel_j_k1}) and (\ref{eq:Rec_rel_j_k}), one
has
\begin{align}
E_{\mu_{3}}^{\left(1,3,2\right)} & =\frac{1}{2}\sum_{\underset{\left(k\ne2\right)}{k=2}}^{3}\xi_{\mu_{k}}E^{\left(k-1,2,1\right)}-\frac{1}{2}\xi^{\mu_{2}}\left(\prod_{i=2}^{2}\nabla_{\mu_{i}}\right)\xi_{\mu_{3}}\nonumber \\
 & =\frac{1}{2}\xi_{\mu_{3}}E^{\left(2,2,1\right)}-\frac{1}{2}\xi^{\mu_{2}}\nabla_{\mu_{2}}\xi_{\mu_{3}},
\end{align}
and
\begin{equation}
E_{\mu_{1}}^{\left(3,3,2\right)}=\nabla_{\mu_{1}}E^{\left(2,2,1\right)}.
\end{equation}
Here, $E^{\left(2,2,1\right)}$ should be obtained from $\lambda^{\mu_{1}}\nabla_{\mu_{1}}\xi_{\mu_{2}}$
and
\begin{equation}
\lambda^{\mu_{1}}\nabla_{\mu_{1}}\xi_{\mu_{2}}=\sum_{\underset{\left(k\ne1\right)}{k=1}}^{2}\lambda_{\mu_{k}}E^{\left(k,2,1\right)}=\lambda_{\mu_{2}}E^{\left(2,2,1\right)},
\end{equation}
and we know that $\lambda^{\mu_{1}}\nabla_{\mu_{1}}\xi_{\mu_{2}}$
satisfies}}
\begin{equation}
\lambda^{\mu_{1}}\nabla_{\mu_{1}}\xi_{\mu_{2}}=-\lambda_{\mu_{2}}\left(\frac{1}{4}\xi^{\mu_{1}}\xi_{\mu_{1}}-\frac{R}{D\left(D-1\right)}\right),
\end{equation}
\textit{\emph{from which $E^{\left(2,2,1\right)}$ can be obtained
as
\begin{equation}
E^{\left(2,2,1\right)}=-\left(\frac{1}{4}\xi^{\mu_{1}}\xi_{\mu_{1}}-\frac{R}{D\left(D-1\right)}\right).
\end{equation}
Using this result, $E_{\mu_{3}}^{\left(1,3,2\right)}$ and $E_{\mu_{1}}^{\left(3,3,2\right)}$
become
\begin{equation}
E_{\mu_{3}}^{\left(1,3,2\right)}=-\frac{1}{2}\xi_{\mu_{3}}\left(\frac{1}{4}\xi^{\mu_{2}}\xi_{\mu_{2}}-\frac{R}{D\left(D-1\right)}\right)-\frac{1}{2}\xi^{\mu_{2}}\nabla_{\mu_{2}}\xi_{\mu_{3}},
\end{equation}
and
\begin{equation}
E_{\mu_{1}}^{\left(3,3,2\right)}=-\frac{1}{2}\xi^{\mu_{2}}\nabla_{\mu_{1}}\xi_{\mu_{2}}.
\end{equation}
With these results, $\lambda^{\mu_{2}}\nabla_{\mu_{1}}\nabla_{\mu_{2}}\xi_{\mu_{3}}$
becomes
\begin{equation}
\lambda^{\mu_{2}}\nabla_{\mu_{1}}\nabla_{\mu_{2}}\xi_{\mu_{3}}=\lambda_{\mu_{1}}\left[-\frac{1}{2}\xi_{\mu_{3}}\left(\frac{1}{4}\xi^{\mu_{2}}\xi_{\mu_{2}}-\frac{R}{D\left(D-1\right)}\right)-\frac{1}{2}\xi^{\mu_{2}}\nabla_{\mu_{2}}\xi_{\mu_{3}}\right]-\frac{1}{2}\lambda_{\mu_{3}}\xi^{\mu_{2}}\nabla_{\mu_{1}}\xi_{\mu_{2}},
\end{equation}
which is the same as (\ref{eq:O3_j2}). Let us also apply the recursive
relations (\ref{eq:Rec_rel_j_k1}) and (\ref{eq:Rec_rel_j_k}) in
the case of the $j=3$ contraction pattern of $\nabla\nabla\xi$ for
which one has
\begin{align}
\lambda^{\mu_{3}}\left(\prod_{i=1}^{2}\nabla_{\mu_{i}}\right)\xi_{\mu_{3}} & =\sum_{\underset{\left(k\ne3\right)}{k=1}}^{3}\lambda_{\mu_{k}}E_{\mu_{r_{_{1}}}}^{\left(k,3,3\right)}=\lambda_{\mu_{1}}E_{\mu_{2}}^{\left(1,3,3\right)}+\lambda_{\mu_{2}}E_{\mu_{1}}^{\left(2,3,3\right)},
\end{align}
and from the recursive relations, one has
\begin{align}
E_{\mu_{2}}^{\left(1,3,3\right)} & =\frac{1}{2}\sum_{\underset{\left(k\ne3\right)}{k=2}}^{3}\xi_{\mu_{k}}E^{\left(k-1,2,2\right)}-\frac{1}{2}\xi^{\mu_{3}}\left(\prod_{i=2}^{2}\nabla_{\mu_{i}}\right)\xi_{\mu_{3}}\nonumber \\
 & =\frac{1}{2}\xi_{\mu_{2}}E^{\left(1,2,2\right)}-\frac{1}{2}\xi^{\mu_{3}}\nabla_{\mu_{2}}\xi_{\mu_{3}},
\end{align}
and
\begin{equation}
E_{\mu_{1}}^{\left(2,3,3\right)}=\nabla_{\mu_{1}}E^{\left(1,2,2\right)}.
\end{equation}
Here, $E^{\left(1,2,2\right)}$ should be obtained from $\lambda^{\mu_{2}}\nabla_{\mu_{1}}\xi_{\mu_{2}}$
and
\begin{equation}
\lambda^{\mu_{2}}\nabla_{\mu_{1}}\xi_{\mu_{2}}=\sum_{\underset{\left(k\ne2\right)}{k=1}}^{2}\lambda_{\mu_{k}}E^{\left(k,2,2\right)}=\lambda_{\mu_{1}}E^{\left(1,2,2\right)},
\end{equation}
and we know that $\lambda^{\mu_{2}}\nabla_{\mu_{1}}\xi_{\mu_{2}}$
satisfies}}
\begin{equation}
\lambda^{\mu_{2}}\nabla_{\mu_{1}}\xi_{\mu_{2}}=-\frac{1}{2}\lambda_{\mu_{1}}\xi^{\mu_{2}}\xi_{\mu_{2}},
\end{equation}
\textit{\emph{from which $E^{\left(1,2,2\right)}$ can be obtained
as
\begin{equation}
E^{\left(1,2,2\right)}=-\frac{1}{2}\xi^{\mu_{2}}\xi_{\mu_{2}}.
\end{equation}
Using this result, $E_{\mu_{2}}^{\left(1,3,3\right)}$ and $E_{\mu_{1}}^{\left(2,3,3\right)}$
become
\begin{equation}
E_{\mu_{2}}^{\left(1,3,3\right)}=-\frac{1}{4}\xi_{\mu_{2}}\xi^{\mu_{3}}\xi_{\mu_{3}}-\frac{1}{2}\xi^{\mu_{3}}\nabla_{\mu_{2}}\xi_{\mu_{3}},
\end{equation}
and
\begin{equation}
E_{\mu_{1}}^{\left(2,3,3\right)}=-\xi^{\mu_{3}}\nabla_{\mu_{1}}\xi_{\mu_{3}}.
\end{equation}
With these results, $\lambda^{\mu_{3}}\nabla_{\mu_{1}}\nabla_{\mu_{2}}\xi_{\mu_{3}}$
becomes
\begin{equation}
\lambda^{\mu_{3}}\nabla_{\mu_{1}}\nabla_{\mu_{2}}\xi_{\mu_{3}}=\lambda_{\mu_{1}}\left(-\frac{1}{4}\xi_{\mu_{2}}\xi^{\mu_{3}}\xi_{\mu_{3}}-\frac{1}{2}\xi^{\mu_{3}}\nabla_{\mu_{2}}\xi_{\mu_{3}}\right)-\lambda_{\mu_{2}}\xi^{\mu_{3}}\nabla_{\mu_{1}}\xi_{\mu_{3}},
\end{equation}
which is the same as (\ref{eq:O3_j3}).}}

\noindent \textbf{\vspace{0.2cm}
}

\noindent \textbf{\textit{\emph{Example 4: }}}\textit{\emph{Now, let
us calculate $\lambda^{\mu_{1}}\nabla_{\mu_{1}}\nabla_{\mu_{2}}\nabla_{\mu_{3}}\xi_{\mu_{4}}$
explicitly and also compute it with the recursion relations (\ref{eq:Rec_rel_j1_k2},\ref{eq:Rec_rel_j1_k},\ref{eq:Rec_rel_j1_kn}).
This calculation demonstrates the usefulness of these recursion relations
at the first nontrivial order. Thus, $\lambda^{\mu_{1}}\nabla_{\mu_{1}}\nabla_{\mu_{2}}\nabla_{\mu_{3}}\xi_{\mu_{4}}$
can be calculated in terms of $\lambda^{\mu_{1}}\nabla_{\mu_{2}}\nabla_{\mu_{1}}\nabla_{\mu_{3}}\xi_{\mu_{4}}$
as
\begin{align}
\lambda^{\mu_{1}}\nabla_{\mu_{1}}\nabla_{\mu_{2}}\nabla_{\mu_{3}}\xi_{\mu_{4}}= & \frac{R}{D\left(D-1\right)}\left(\lambda_{\mu_{3}}\nabla_{\mu_{2}}\xi_{\mu_{4}}+\lambda_{\mu_{4}}\nabla_{\mu_{3}}\xi_{\mu_{2}}\right)\nonumber \\
 & -\frac{R}{D\left(D-1\right)}\left(g_{\mu_{2}\mu_{3}}\lambda^{\mu_{5}}\nabla_{\mu_{5}}\xi_{\mu_{4}}+g_{\mu_{2}\mu_{4}}\lambda^{\mu_{5}}\nabla_{\mu_{3}}\xi_{\mu_{5}}\right)\nonumber \\
 & +\lambda^{\mu_{1}}\nabla_{\mu_{2}}\nabla_{\mu_{1}}\nabla_{\mu_{3}}\xi_{\mu_{4}}.
\end{align}
Then, calculating $\lambda^{\mu_{1}}\nabla_{\mu_{2}}\nabla_{\mu_{1}}\nabla_{\mu_{3}}\xi_{\mu_{4}}$
yields
\begin{align}
\lambda^{\mu_{1}}\nabla_{\mu_{2}}\nabla_{\mu_{1}}\nabla_{\mu_{3}}\xi_{\mu_{4}}= & -\frac{1}{2}\lambda_{\mu_{2}}\left[\xi^{\mu_{1}}\nabla_{\mu_{1}}\nabla_{\mu_{3}}\xi_{\mu_{4}}+\frac{1}{2}\xi_{\mu_{3}}\xi^{\mu_{1}}\nabla_{\mu_{1}}\xi_{\mu_{4}}+\frac{1}{2}\xi_{\mu_{4}}\xi^{\mu_{1}}\nabla_{\mu_{3}}\xi_{\mu_{1}}\right]\nonumber \\
 & -\frac{1}{4}\lambda_{\mu_{2}}\xi_{\mu_{3}}\xi_{\mu_{4}}\left(\frac{1}{4}\xi^{\mu_{1}}\xi_{\mu_{1}}-\frac{3R}{D\left(D-1\right)}\right)\nonumber \\
 & -\frac{1}{2}\lambda_{\mu_{3}}\left[\xi^{\mu_{1}}\nabla_{\mu_{2}}\nabla_{\mu_{1}}\xi_{\mu_{4}}+\left(\nabla_{\mu_{2}}\xi^{\mu_{1}}\right)\nabla_{\mu_{1}}\xi_{\mu_{4}}\right]\nonumber \\
 & -\frac{1}{2}\lambda_{\mu_{3}}\left[\left(\nabla_{\mu_{2}}\xi_{\mu_{4}}\right)\left(\frac{1}{4}\xi^{\mu_{1}}\xi_{\mu_{1}}-\frac{R}{D\left(D-1\right)}\right)+\frac{1}{2}\xi_{\mu_{4}}\xi^{\mu_{1}}\nabla_{\mu_{2}}\xi_{\mu_{1}}\right]\nonumber \\
 & -\frac{1}{2}\lambda_{\mu_{4}}\left[\xi^{\mu_{1}}\nabla_{\mu_{2}}\nabla_{\mu_{3}}\xi_{\mu_{1}}+\left(\nabla_{\mu_{2}}\xi^{\mu_{1}}\right)\nabla_{\mu_{3}}\xi_{\mu_{1}}-\frac{2R}{D\left(D-1\right)}\nabla_{\mu_{2}}\xi_{\mu_{3}}\right].
\end{align}
Using this result, $\lambda^{\mu_{1}}\nabla_{\mu_{1}}\nabla_{\mu_{2}}\nabla_{\mu_{3}}\xi_{\mu_{4}}$
becomes
\begin{align}
\lambda^{\mu_{1}}\nabla_{\mu_{1}}\nabla_{\mu_{2}}\nabla_{\mu_{3}}\xi_{\mu_{4}}= & -\frac{1}{2}\lambda_{\mu_{2}}\left[\xi^{\mu_{1}}\nabla_{\mu_{1}}\nabla_{\mu_{3}}\xi_{\mu_{4}}+\frac{1}{2}\xi_{\mu_{3}}\xi^{\mu_{1}}\nabla_{\mu_{1}}\xi_{\mu_{4}}+\frac{1}{2}\xi_{\mu_{4}}\xi^{\mu_{1}}\nabla_{\mu_{3}}\xi_{\mu_{1}}\right]\nonumber \\
 & -\frac{1}{4}\lambda_{\mu_{2}}\xi_{\mu_{3}}\xi_{\mu_{4}}\left(\frac{1}{4}\xi^{\mu_{1}}\xi_{\mu_{1}}-\frac{3R}{D\left(D-1\right)}\right)\nonumber \\
 & -\frac{1}{2}\lambda_{\mu_{3}}\left[\xi^{\mu_{1}}\nabla_{\mu_{2}}\nabla_{\mu_{1}}\xi_{\mu_{4}}+\left(\nabla_{\mu_{2}}\xi^{\mu_{1}}\right)\nabla_{\mu_{1}}\xi_{\mu_{4}}-\frac{R}{D\left(D-1\right)}g_{\mu_{2}\mu_{4}}\xi^{\mu_{1}}\xi_{\mu_{1}}\right]\nonumber \\
 & -\frac{1}{2}\lambda_{\mu_{3}}\left[\left(\nabla_{\mu_{2}}\xi_{\mu_{4}}\right)\left(\frac{1}{4}\xi^{\mu_{1}}\xi_{\mu_{1}}-\frac{3R}{D\left(D-1\right)}\right)+\frac{1}{2}\xi_{\mu_{4}}\xi^{\mu_{1}}\nabla_{\mu_{2}}\xi_{\mu_{1}}\right]\nonumber \\
 & +\frac{R}{D\left(D-1\right)}\lambda_{\mu_{4}}\left[g_{\mu_{2}\mu_{3}}\left(\frac{1}{4}\xi^{\mu_{1}}\xi_{\mu_{1}}-\frac{R}{D\left(D-1\right)}\right)+\nabla_{\mu_{3}}\xi_{\mu_{2}}+\nabla_{\mu_{2}}\xi_{\mu_{3}}\right]\nonumber \\
 & -\frac{1}{2}\lambda_{\mu_{4}}\left[\xi^{\mu_{1}}\nabla_{\mu_{2}}\nabla_{\mu_{3}}\xi_{\mu_{1}}+\left(\nabla_{\mu_{2}}\xi^{\mu_{1}}\right)\nabla_{\mu_{3}}\xi_{\mu_{1}}\right].\label{eq:O4_j1}
\end{align}
Now, let us find this result from the recursion relations. The $E_{\mu_{r_{_{1}}}\mu_{r_{_{2}}}\dots\mu_{r_{_{n-2}}}}^{\left(k,n,1\right)}$
terms that we need to calculate are
\begin{align}
\lambda^{\mu_{1}}\left(\prod_{i=1}^{3}\nabla_{\mu_{i}}\right)\xi_{\mu_{4}} & =\sum_{k=2}^{4}\lambda_{\mu_{k}}E_{\mu_{r_{_{1}}}\mu_{r_{_{2}}}}^{\left(k,4,1\right)}\nonumber \\
\lambda^{\mu_{1}}\nabla_{\mu_{1}}\nabla_{\mu_{2}}\nabla_{\mu_{3}}\xi_{\mu_{4}} & =\lambda_{\mu_{2}}E_{\mu_{3}\mu_{4}}^{\left(2,4,1\right)}+\lambda_{\mu_{3}}E_{\mu_{2}\mu_{4}}^{\left(3,4,1\right)}+\lambda_{\mu_{4}}E_{\mu_{2}\mu_{3}}^{\left(4,4,1\right)}.
\end{align}
Using the recursion relations (\ref{eq:Rec_rel_j1_k2},\ref{eq:Rec_rel_j1_k},\ref{eq:Rec_rel_j1_kn}),
one has
\begin{align}
E_{\mu_{3}\mu_{4}}^{\left(2,4,1\right)} & =\frac{1}{2}\left[\sum_{k=3}^{4}\xi_{\mu_{k}}E_{\mu_{m_{_{1}}}}^{\left(k-1,3,1\right)}-\xi^{\mu_{1}}\nabla_{\mu_{1}}\left(\prod_{i=3}^{3}\nabla_{\mu_{i}}\right)\xi_{\mu_{4}}\right]\nonumber \\
 & =\frac{1}{2}\left[\xi_{\mu_{3}}E_{\mu_{4}}^{\left(2,3,1\right)}+\xi_{\mu_{4}}E_{\mu_{3}}^{\left(3,3,1\right)}-\xi^{\mu_{1}}\nabla_{\mu_{1}}\nabla_{\mu_{3}}\xi_{\mu_{4}}\right],
\end{align}
\begin{align}
E_{\mu_{2}\mu_{4}}^{\left(3,4,1\right)}= & \frac{R}{D\left(D-1\right)}\left(\prod_{i_{1}=3}^{2}\nabla_{\mu_{i_{_{1}}}}\right)\nabla_{\mu_{2}}\left(\prod_{i_{2}=4}^{3}\nabla_{\mu_{i_{_{2}}}}\right)\xi_{\mu_{4}}\nonumber \\
 & -\frac{R}{D\left(D-1\right)}\sum_{\underset{\left(s\ne3\right)}{s=3}}^{4}g_{\mu_{2}\mu_{s}}E^{\left(1,2,s-2\right)}+\nabla_{\mu_{2}}E_{\mu_{4}}^{\left(2,3,1\right)}\nonumber \\
= & \frac{R}{D\left(D-1\right)}\left(\nabla_{\mu_{2}}\xi_{\mu_{4}}-g_{\mu_{2}\mu_{4}}E^{\left(1,2,2\right)}\right)+\nabla_{\mu_{2}}E_{\mu_{4}}^{\left(2,3,1\right)},
\end{align}
\begin{align}
E_{\mu_{2}\mu_{3}}^{\left(4,4,1\right)} & =\frac{R}{D\left(D-1\right)}\left[\left(\prod_{i=3}^{3}\nabla_{\mu_{i}}\right)\xi_{\mu_{2}}-\sum_{s=3}^{3}g_{\mu_{2}\mu_{s}}E^{\left(2,2,s-2\right)}\right]+\nabla_{\mu_{2}}E_{\mu_{3}}^{\left(3,3,1\right)}\nonumber \\
 & =\frac{R}{D\left(D-1\right)}\left(\nabla_{\mu_{3}}\xi_{\mu_{2}}-g_{\mu_{2}\mu_{3}}E^{\left(2,2,1\right)}\right)+\nabla_{\mu_{2}}E_{\mu_{3}}^{\left(3,3,1\right)}.
\end{align}
We have already calculated $E^{\left(1,2,2\right)}$ and $E^{\left(2,2,1\right)}$.
The term $E_{\mu_{4}}^{\left(2,3,1\right)}$ is the coefficient of
$\lambda_{\mu_{2}}$ in $\lambda^{\mu_{1}}\nabla_{\mu_{1}}\nabla_{\mu_{2}}\xi_{\mu_{3}}$
which is
\begin{equation}
E_{\mu_{4}}^{\left(2,3,1\right)}=-\frac{1}{2}\left[\xi_{\mu_{4}}\left(\frac{1}{4}\xi^{\mu_{1}}\xi_{\mu_{1}}-\frac{R}{D\left(D-1\right)}\right)+\xi^{\mu_{1}}\nabla_{\mu_{1}}\xi_{\mu_{4}}\right].
\end{equation}
The term $E_{\mu_{3}}^{\left(3,3,1\right)}$ is the coefficient of
$\lambda_{\mu_{3}}$ in, again, $\lambda^{\mu_{1}}\nabla_{\mu_{1}}\nabla_{\mu_{2}}\xi_{\mu_{3}}$
which is
\begin{equation}
E_{\mu_{3}}^{\left(3,3,1\right)}=-\frac{1}{2}\xi^{\mu_{1}}\nabla_{\mu_{3}}\xi_{\mu_{1}}+\frac{R}{D\left(D-1\right)}\xi_{\mu_{3}}.
\end{equation}
Putting these results in $E_{\mu_{3}\mu_{4}}^{\left(2,4,1\right)}$,
$E_{\mu_{2}\mu_{4}}^{\left(3,4,1\right)}$, and $E_{\mu_{2}\mu_{3}}^{\left(4,4,1\right)}$
yields
\begin{align}
E_{\mu_{3}\mu_{4}}^{\left(2,4,1\right)}= & -\frac{1}{4}\xi_{\mu_{3}}\xi_{\mu_{4}}\left(\frac{1}{4}\xi^{\mu_{1}}\xi_{\mu_{1}}-\frac{3R}{D\left(D-1\right)}\right)\nonumber \\
 & -\frac{1}{2}\left(\xi^{\mu_{1}}\nabla_{\mu_{1}}\nabla_{\mu_{3}}\xi_{\mu_{4}}+\frac{1}{2}\xi_{\mu_{4}}\xi^{\mu_{1}}\nabla_{\mu_{3}}\xi_{\mu_{1}}+\frac{1}{2}\xi_{\mu_{3}}\xi^{\mu_{1}}\nabla_{\mu_{1}}\xi_{\mu_{4}}\right),
\end{align}
\begin{align}
E_{\mu_{2}\mu_{4}}^{\left(3,4,1\right)}= & -\frac{1}{2}\left[\left(\nabla_{\mu_{2}}\xi_{\mu_{4}}\right)\left(\frac{1}{4}\xi^{\mu_{1}}\xi_{\mu_{1}}-\frac{3R}{D\left(D-1\right)}\right)+\frac{1}{2}\xi_{\mu_{4}}\xi^{\mu_{1}}\nabla_{\mu_{2}}\xi_{\mu_{1}}\right]\nonumber \\
 & -\frac{1}{2}\left[\xi^{\mu_{1}}\nabla_{\mu_{2}}\nabla_{\mu_{1}}\xi_{\mu_{4}}+\left(\nabla_{\mu_{2}}\xi^{\mu_{1}}\right)\nabla_{\mu_{1}}\xi_{\mu_{4}}-\frac{R}{D\left(D-1\right)}g_{\mu_{2}\mu_{4}}\xi^{\mu_{1}}\xi_{\mu_{1}}\right],
\end{align}
\begin{align}
E_{\mu_{2}\mu_{3}}^{\left(4,4,1\right)}= & \frac{R}{D\left(D-1\right)}\left[\nabla_{\mu_{2}}\xi_{\mu_{3}}+\nabla_{\mu_{3}}\xi_{\mu_{2}}+g_{\mu_{2}\mu_{3}}\left(\frac{1}{4}\xi^{\mu_{1}}\xi_{\mu_{1}}-\frac{R}{D\left(D-1\right)}\right)\right]\nonumber \\
 & -\frac{1}{2}\left[\xi^{\mu_{1}}\nabla_{\mu_{2}}\nabla_{\mu_{3}}\xi_{\mu_{1}}+\left(\nabla_{\mu_{2}}\xi^{\mu_{1}}\right)\nabla_{\mu_{3}}\xi_{\mu_{1}}\right],
\end{align}
which are the same as the ones that can be obtained from (\ref{eq:O4_j1}).
After these lemmas and examples, we now have the proper arsenal to
prove the theorem. }}\textbf{\vspace{0.5cm}
}

\noindent \textbf{\textit{\emph{Proof of Theorem 2: }}}\textit{\emph{As
a result of the previous lemmas, we showed that $\lambda^{\mu_{j}}\left(\prod_{i=1}^{n-1}\nabla_{\mu_{i}}\right)\xi_{\mu_{n}}$
is $\lambda$-reducible as
\[
\lambda^{\mu_{j}}\left(\prod_{i=1}^{n-1}\nabla_{\mu_{i}}\right)\xi_{\mu_{n}}=\sum_{\underset{\left(k\ne j\right)}{k=1}}^{n}\lambda_{\mu_{k}}E_{\mu_{r_{_{1}}}\mu_{r_{_{2}}}\dots\mu_{r_{_{n-2}}}}^{\left(k,n,j\right)},
\]
where the $\left(0,n-2\right)$ rank tensors $E_{\mu_{r_{_{1}}}\mu_{r_{_{2}}}\dots\mu_{r_{_{n-2}}}}^{\left(k,n,j\right)}$
are related to the lower orders with the recursion relations (\ref{eq:Rec_rel_j1_k2},\ref{eq:Rec_rel_j1_k},\ref{eq:Rec_rel_j1_kn})
for the $j=1$ contraction pattern and with the recursion relations
(\ref{eq:Rec_rel_j_k1},\ref{eq:Rec_rel_j_k}) for the contraction
patterns of $1<j\le n$. From these recursion relations, one can see
that the $E_{\mu_{r_{_{1}}}\mu_{r_{_{2}}}\dots\mu_{r_{_{n-2}}}}^{\left(k,n,j\right)}$
tensors are built from the structures
\begin{equation}
\xi_{\mu_{k}},\qquad E_{\mu_{m_{_{1}}}\mu_{m_{_{2}}}\dots\mu_{m_{_{n-3}}}}^{\left(k,n-1,j\right)},\qquad\left(\prod_{i=1}^{n-2}\nabla_{\mu_{i}}\right)\xi_{\mu_{n-1}},\qquad\nabla_{\mu_{1}}E_{\mu_{m_{_{1}}}\mu_{m_{_{2}}}\dots\mu_{m_{_{n-3}}}}^{\left(k,n-1,j\right)},
\end{equation}
\begin{equation}
g_{\mu_{2}\mu_{s}},\qquad E_{\mu_{m_{_{1}}}\mu_{m_{_{2}}}\dots\mu_{m_{_{n-4}}}}^{\left(k,n-2,j\right)},\qquad\left(\prod_{i=1}^{n-3}\nabla_{\mu_{i}}\right)\xi_{\mu_{n-2}},
\end{equation}
where the building blocks of $E_{\mu_{m_{_{1}}}\mu_{m_{_{2}}}\dots\mu_{m_{_{n-3}}}}^{\left(k,n-1,j\right)}$
and $E_{\mu_{m_{_{1}}}\mu_{m_{_{2}}}\dots\mu_{m_{_{n-4}}}}^{\left(k,n-2,j\right)}$
are
\begin{equation}
g_{\mu_{1}\mu_{2}},\qquad\xi_{\mu_{1}},\qquad\left(\prod_{i=1}^{r-1}\nabla_{\mu_{i}}\right)\xi_{\mu_{r}},\qquad r=2,3,\dots,n-2,
\end{equation}
and
\begin{equation}
g_{\mu_{1}\mu_{2}},\qquad\xi_{\mu_{1}},\qquad\left(\prod_{i=1}^{r-1}\nabla_{\mu_{i}}\right)\xi_{\mu_{r}},\qquad r=2,3,\dots,n-3,
\end{equation}
respectively. Therefore, the building blocks of the $E_{\mu_{r_{_{1}}}\mu_{r_{_{2}}}\dots\mu_{r_{_{n-2}}}}^{\left(k,n,j\right)}$
tensors are
\begin{equation}
g_{\mu_{1}\mu_{2}},\qquad\xi_{\mu_{1}},\qquad\left(\prod_{i=1}^{r-1}\nabla_{\mu_{i}}\right)\xi_{\mu_{r}},\qquad r=2,3,\dots,n-1.
\end{equation}
Note that contracting the $\left(0,n\right)$ rank tensor $\nabla^{n-1}\xi$
with $\lambda$ reduces the derivative order such that the highest
derivative order term now becomes $\nabla^{n-2}\xi$.}}

\textit{\emph{Now, let us discuss the contractions of $\nabla^{n-1}\xi$
with more than one $\lambda$. In this regard, the important thing
that should be noticed in the one-$\lambda$ contraction result is
that the building blocks of the tensor structures produced by the
one-$\lambda$ contraction of $\nabla^{n-1}\xi$, which are the metric,
$\xi$, and the lower order derivatives of $\xi$, are all $\lambda$-reducible
under one-$\lambda$ contraction. For a further $\lambda$ contraction,
when the $\lambda$ vector is contracted with the derivatives of $\xi$,
again the lower order derivatives of $\xi$ will appear together with
the metric and $\xi$ as building blocks. Due to continuous appearance
of the same $\lambda$-reducible building blocks, $\nabla^{n-1}\xi$
should be $\lambda$-conserving. }}

\textit{\emph{To be more explicit, let us first consider the contraction
of the $\left(0,n\right)$ rank tensor $\nabla^{n-1}\xi$ with two
$\lambda$ vectors; that is
\begin{equation}
\lambda^{\mu_{j_{_{1}}}}\lambda^{\mu_{j_{_{2}}}}\left(\prod_{i=1}^{n-1}\nabla_{\mu_{i}}\right)\xi_{\mu_{n}},
\end{equation}
where $\left\{ j_{1},j_{2}\right\} $ is a subset of $\left\{ 1,2,\dots n\right\} $.
Using the one-$\lambda$ contraction result (\ref{eq:One_l_cont_On}),
one has
\begin{equation}
\lambda^{\mu_{j_{_{1}}}}\lambda^{\mu_{j_{_{2}}}}\left(\prod_{i=1}^{n-1}\nabla_{\mu_{i}}\right)\xi_{\mu_{n}}=\sum_{\underset{\left(k_{1}\ne j_{1},j_{2}\right)}{k_{1}=1}}^{n}\lambda_{\mu_{k_{_{1}}}}\lambda^{\mu_{j_{_{2}}}}E_{\mu_{m_{_{1}}}\mu_{m_{_{2}}}\dots\mu_{m_{_{n-2}}}}^{\left(k_{1},n,j_{1}\right)},
\end{equation}
where we know that the $E_{\mu_{m_{_{1}}}\mu_{m_{_{2}}}\dots\mu_{m_{_{n-2}}}}^{\left(k_{1},n,j_{1}\right)}$
tensors are the rank $\left(0,n-2\right)$ contractions of the building
blocks
\begin{equation}
g_{\mu_{1}\mu_{2}},\qquad\xi_{\mu_{1}},\qquad\left(\prod_{i=1}^{r-1}\nabla_{\mu_{i}}\right)\xi_{\mu_{r}},\qquad r=2,3,\dots,n-1.\label{eq:Building_blocks_of_one_l_En}
\end{equation}
Since the one-$\lambda$ contraction of these building blocks are
$\lambda$-reducible, the $E_{\mu_{m_{_{1}}}\mu_{m_{_{2}}}\dots\mu_{m_{_{n-2}}}}^{\left(k_{1},n,j_{1}\right)}$
tensors should also be $\lambda$-reducible as
\begin{equation}
\lambda^{\mu_{j_{_{2}}}}E_{\mu_{m_{_{1}}}\mu_{m_{_{2}}}\dots\mu_{m_{_{n-2}}}}^{\left(k_{1},n,j_{1}\right)}=\sum_{\underset{\left(k_{2}\ne k_{1},j_{1},j_{2}\right)}{k_{2}=1}}^{n}\lambda_{\mu_{_{k_{2}}}}E_{\mu_{r_{_{1}}}\mu_{r_{_{2}}}\dots\mu_{r_{_{n-4}}}}^{\left(k_{1},k_{2},n,j_{1},j_{2}\right)},
\end{equation}
where $\left(r_{1},r_{2},\dots,r_{n-4}\right)$ is an }}\textit{increasing}\textit{\emph{
sequence constructed from $\left\{ 1,2,\dots,n\right\} \backslash\left\{ j_{1},k_{1},j_{2},k_{2}\right\} $.
We know that contracting $\nabla^{n-1}\xi$ with one $\lambda$ reduces
the highest derivative order to $\nabla^{n-2}\xi$. Then, the highest
derivative order for the building blocks of $E_{\mu_{r_{_{1}}}\mu_{r_{_{2}}}\dots\mu_{r_{_{n-4}}}}^{\left(k_{1},k_{2},n,j_{1},j_{2}\right)}$
should be one order less than the highest derivative order for the
building blocks of $E_{\mu_{m_{_{1}}}\mu_{m_{_{2}}}\dots\mu_{m_{_{n-2}}}}^{\left(k_{1},n,j_{1}\right)}$
given in (\ref{eq:Building_blocks_of_one_l_En}). Thus, the $E_{\mu_{r_{_{1}}}\mu_{r_{_{2}}}\dots\mu_{r_{_{n-4}}}}^{\left(k_{1},k_{2},n,j_{1},j_{2}\right)}$
tensors are the rank $\left(0,n-4\right)$ contractions of the building
blocks
\begin{equation}
g_{\mu_{1}\mu_{2}},\qquad\xi_{\mu_{1}},\qquad\left(\prod_{i=1}^{r-1}\nabla_{\mu_{i}}\right)\xi_{\mu_{r}},\qquad r=2,3,\dots,n-2.
\end{equation}
The final form of the two-$\lambda$ contraction of $\nabla^{n-1}\xi$,
\begin{equation}
\lambda^{\mu_{j_{_{1}}}}\lambda^{\mu_{j_{_{2}}}}\left(\prod_{i=1}^{n-1}\nabla_{\mu_{i}}\right)\xi_{\mu_{n}},
\end{equation}
becomes
\begin{align}
\lambda^{\mu_{j_{_{1}}}}\lambda^{\mu_{j_{_{2}}}}\left(\prod_{i=1}^{n-1}\nabla_{\mu_{i}}\right)\xi_{\mu_{n}}= & \sum_{\underset{\left(k_{1}\ne j_{1},j_{2}\right)}{k_{1}=1}}^{n}\lambda_{\mu_{k_{_{1}}}}\sum_{\underset{\left(k_{2}\ne k_{1},j_{1},j_{2}\right)}{k_{2}=1}}^{n}\lambda_{\mu_{_{k_{2}}}}E_{\mu_{r_{_{1}}}\mu_{r_{_{2}}}\dots\mu_{r_{_{n-4}}}}^{\left(k_{1},k_{2},n,j_{1},j_{2}\right)}\nonumber \\
= & \sum_{\underset{\left(k_{1}\ne j_{1},j_{2}\right)}{k_{1}=1}}^{n}\sum_{\underset{\left(k_{2}\ne k_{1},j_{1},j_{2}\right)}{k_{2}=1}}^{n}\lambda_{\mu_{k_{_{1}}}}\lambda_{\mu_{_{k_{2}}}}E_{\mu_{r_{_{1}}}\mu_{r_{_{2}}}\dots\mu_{r_{_{n-4}}}}^{\left(k_{1},k_{2},n,j_{1},j_{2}\right)}.
\end{align}
Here, notice the pattern that the two-$\lambda$ contraction of $\nabla^{n-1}\xi$
becomes a sum of $\left(0,n-2\right)$ tensors which are decomposed
into two $\lambda$ vectors and rank $\left(0,n-4\right)$ tensors
$E_{\mu_{r_{_{1}}}\mu_{r_{_{2}}}\dots\mu_{r_{_{n-4}}}}^{\left(k_{1},k_{2},n,j_{1},j_{2}\right)}$
while the one-$\lambda$ contraction of $\nabla^{n-1}\xi$ becomes
a sum of $\left(0,n-1\right)$ tensors which are decomposed into one
$\lambda$ vector and rank $\left(0,n-2\right)$ tensors $E_{\mu_{r_{_{1}}}\mu_{r_{_{2}}}\dots\mu_{r_{_{n-2}}}}^{\left(k_{1},n,j_{1}\right)}$.}}

\textit{\emph{Further $\lambda$ contractions of $\nabla^{n-1}\xi$
also have the same pattern: contracting the rank $\left(0,n\right)$
tensor $\nabla^{n-1}\xi$ with $p$ number of $\lambda$ tensors yields
a sum of $\left(0,n-p\right)$ rank tensors which can be decomposed
into $p$ number of $\lambda$ tensors and rank $\left(0,n-2p\right)$
tensors
\begin{equation}
E_{\mu_{r_{_{1}}}\mu_{r_{_{2}}}\dots\mu_{r_{_{n-2p}}}}^{\left(k_{1},k_{2},\dots,k_{p},n,j_{1},j_{2},\dots,j_{p}\right)},
\end{equation}
for which the building blocks are
\begin{equation}
g_{\mu_{1}\mu_{2}},\qquad\xi_{\mu_{1}},\qquad\left(\prod_{i=1}^{r-1}\nabla_{\mu_{i}}\right)\xi_{\mu_{r}},\qquad r=2,3,\dots,n-p.
\end{equation}
More explicitly, the contraction of $\nabla^{n-1}\xi$ with $p$ number
of $\lambda$ vectors can be represented as
\begin{equation}
\left(\prod_{r=1}^{p}\lambda^{\mu_{j_{_{r}}}}\right)\left(\prod_{i=1}^{n-1}\nabla_{\mu_{i}}\right)\xi_{\mu_{n}},
\end{equation}
where $\left\{ j_{1},j_{2},\dots,j_{p}\right\} $ is a subset of $\left\{ 1,2,\dots,n\right\} $,
and following the pattern we developed, this term becomes}}%
\footnote{\textit{\emph{Assuming $n$ is sufficiently larger than $p$.}}%
}\textit{\emph{
\begin{equation}
\left(\prod_{s=1}^{p}\lambda^{\mu_{j_{s}}}\right)\left(\prod_{i=1}^{n-1}\nabla_{\mu_{i}}\right)\xi_{\mu_{n}}=\prod_{s=1}^{p}\left(\sum_{\underset{\left(k_{s}\ne j_{1},\dots,j_{p},k_{1},\dots,k_{s-1}\right)}{k_{s}=1}}^{n}\lambda_{\mu_{k_{s}}}\right)E_{\mu_{r_{_{1}}}\mu_{r_{_{2}}}\dots\mu_{r_{_{n-2p}}}}^{\left(k_{1},k_{2},\dots,k_{p},n,j_{1},j_{2},\dots,j_{p}\right)},
\end{equation}
where $\left\{ r_{1},r_{2},\dots,r_{n-2p}\right\} $ is an increasing
sequence constructed from $\left\{ 1,2,\dots,n\right\} \backslash\left\{ j_{i},k_{i}:\,1\le k\le p\right\} $.
This result shows that the maximum number of $\lambda$ contractions
with $\nabla^{n-1}\xi$ before getting a zero is $n/2$ for even $n$
and $\left(n-1\right)/2$ for odd $n$, and one gets the following
results, respectively;
\begin{equation}
\left(\prod_{s=1}^{\frac{n}{2}}\lambda^{\mu_{j_{s}}}\right)\left(\prod_{i=1}^{n-1}\nabla_{\mu_{i}}\right)\xi_{\mu_{n}}=\prod_{s=1}^{\frac{n}{2}}\left(\sum_{\underset{\left(k_{s}\ne j_{1},\dots,j_{n/2},k_{1},\dots,k_{s-1}\right)}{k_{s}=1}}^{n}\lambda_{\mu_{k_{s}}}\right)E^{\left(k_{1},k_{2},\dots,k_{n/2},n,j_{1},j_{2},\dots,j_{n/2}\right)},
\end{equation}
where the building blocks for $E^{\left(k_{1},k_{2},\dots,k_{n/2},n,j_{1},j_{2},\dots,j_{n/2}\right)}$
are
\begin{equation}
\xi_{\mu_{1}},\qquad\left(\prod_{i=1}^{r-1}\nabla_{\mu_{i}}\right)\xi_{\mu_{r}},\qquad r=2,3,\dots,\frac{n}{2},
\end{equation}
and
\begin{equation}
\left(\prod_{s=1}^{\frac{n-1}{2}}\lambda^{\mu_{j_{s}}}\right)\left(\prod_{i=1}^{n-1}\nabla_{\mu_{i}}\right)\xi_{\mu_{n}}=\prod_{s=1}^{\frac{n-1}{2}}\left(\sum_{\underset{\left(k_{s}\ne j_{1},\dots,j_{\left(n-1\right)/2},k_{1},\dots,k_{s-1}\right)}{k_{s}=1}}^{n}\lambda_{\mu_{k_{s}}}\right)E_{\mu_{m_{_{1}}}}^{\left(k_{1},k_{2},\dots,k_{\left(n-1\right)/2},n,j_{1},j_{2},\dots,j_{\left(n-1\right)/2}\right)},
\end{equation}
where $m_{1}\in\left\{ 1,2,\dots,n\right\} \backslash\left\{ j_{i},k_{i}:\,1\le i\le\left(n-1\right)/2\right\} $
and the building blocks for $E_{\mu_{m_{_{1}}}}^{\left(k_{1},k_{2},\dots,k_{\left(n-1\right)/2},n,j_{1},j_{2},\dots,j_{\left(n-1\right)/2}\right)}$
are
\begin{equation}
\xi_{\mu_{1}},\qquad\left(\prod_{i=1}^{r-1}\nabla_{\mu_{i}}\right)\xi_{\mu_{r}},\qquad r=2,3,\dots,\frac{n+1}{2}.
\end{equation}
}}

\textit{\emph{To conclude, the $\left(0,n\right)$ rank tensor $\nabla^{n-1}\xi$
is $\lambda$-conserving since with each $\lambda$ contraction, the
$\lambda$-weight of the resulting tensor structure increases by one.}}
This proves the theorem. \textit{\emph{\QEDA}}

\textbf{\vspace{0.2cm}
}

\noindent \textbf{\textit{\emph{Example 5: }}}\textit{\emph{For odd
$n$, let us consider $n=3$ case, that is $\nabla\nabla\xi$, for
which one $\lambda$ contractions that we found in (\ref{eq:O3_j2},
\ref{eq:O3_j1}, \ref{eq:O3_j3}) are the last nonzero terms. It can
be verified immediately that a further $\lambda$ contraction with
any of the one-$\lambda$ contraction patterns of $\nabla\nabla\xi$
given in (\ref{eq:O3_j2}, \ref{eq:O3_j1}, \ref{eq:O3_j3}) yields
a zero. For the even $n$ case, let us consider $\nabla^{3}\xi$ for
which two $\lambda$ contraction is the last nonzero order. As an
example, let us study a further $\lambda$ contraction of the $j_{1}=1$
contraction pattern of $\nabla^{3}\xi$ given in (\ref{eq:O4_j1}),
so after a long calculation $\lambda^{\mu_{1}}\lambda^{\mu_{2}}\nabla_{\mu_{1}}\nabla_{\mu_{2}}\nabla_{\mu_{3}}\xi_{\mu_{4}}$
reduces to
\begin{align}
\lambda^{\mu_{1}}\lambda^{\mu_{2}}\nabla_{\mu_{1}}\nabla_{\mu_{2}}\nabla_{\mu_{3}}\xi_{\mu_{4}}= & \frac{1}{2}\lambda_{\mu_{3}}\lambda_{\mu_{4}}\xi^{\mu_{1}}\xi^{\mu_{2}}\nabla_{\mu_{1}}\xi_{\mu_{2}}\nonumber \\
 & -\frac{R}{D\left(D-1\right)}\lambda_{\mu_{3}}\lambda_{\mu_{4}}\left(\frac{3}{4}\xi^{\mu_{2}}\xi_{\mu_{2}}-\frac{R}{D\left(D-1\right)}\right),
\end{align}
which is, as expected, the last nonzero order constructed from the
building blocks $\xi_{\mu}$ and $\nabla_{\mu}\xi_{\nu}$.}}

\textbf{\vspace{0.5cm}
}

\noindent \textbf{\textit{\emph{Theorem 3:}}} \textit{The rank $\left(0,n\right)$
tensor $\nabla^{n}V$ is $\lambda$-conserving.} \textbf{\vspace{0.2cm}
}

\noindent \textbf{\textit{\emph{Proof:}}}\textit{\emph{ The proof
follows the same lines as the proof of Theorem 2. For the first step
of the induction part of the proof, now one has the equations $\lambda^{\mu}\partial_{\mu}V=0$
and
\begin{equation}
\lambda^{\mu}\nabla_{\mu}\partial_{\nu}V=\lambda^{\mu}\nabla_{\nu}\partial_{\mu}V=-\frac{1}{2}\lambda_{\nu}\xi^{\mu}\partial_{\mu}V.
\end{equation}
In addition, the tensorial structures now involve the covariant derivatives
of $V$ in addition to $\xi_{\mu}$ and its covariant derivatives.
Since the proof involves the same cumbersome steps without new ideas,
we do not display it here.\QEDA}}

\noindent \textbf{\vspace{0.5cm}
}

Now, we can give the proof of the main theorem in the text (Theorem
1) based on above results:\textbf{\vspace{0.5cm}
}

\noindent \textbf{\textit{\emph{Proof of Theorem 1: }}}Remember that
$\mathcal{E}_{\mu_{1}\dots\mu_{s}}$ represents the sum of rank $\left(0,s\right)$
tensors which can be decomposed into $2\left(n_{0}+m\right)$ number
of $\lambda$ vectors and rank $\left(0,s-2n_{0}-2m\right)$ tensor
structures which are obtained from the contractions of the following
building blocks\textit{\emph{
\begin{equation}
g_{\mu_{1}\mu_{2}},\qquad\xi_{\mu_{1}},\qquad\left(\prod_{i=1}^{r}\nabla_{\mu_{i}}\right)\xi_{\mu_{r+1}},\qquad\left(\prod_{i=1}^{r+2}\nabla_{\mu_{i}}\right)V,\qquad r=1,2,\dots,n_{m},
\end{equation}
which are all $\lambda$-conserving as we have shown. Then, to have
a nonzero $E_{\mu\nu}$ two-tensor out of $\mathcal{E}_{\mu_{1}\dots\mu_{s}}$,
one must have at most two $\lambda$ vectors in $\mathcal{E}_{\mu_{1}\dots\mu_{s}}$.
If there is more than two $\lambda$ vectors in $\mathcal{E}_{\mu_{1}\dots\mu_{s}}$,
then they eventually yield a zero contraction since any nonzero contraction
for $\mathcal{E}_{\mu_{1}\dots\mu_{s}}$ conserves the number of the
$\lambda$ vectors. Thus, one should start with $R_{\alpha\beta\rho\sigma}$
or $\nabla_{\mu_{1}}\nabla_{\mu_{2}}\dots\nabla_{\mu_{m}}R_{\alpha\beta\rho\sigma}$
to have nonzero $E_{\mu\nu}$ two-tensors. The remaining part of the
proof on the structure of the nonzero $E_{\mu\nu}$ tensors follows
as given in Sec.~\ref{sec:Universality-Of-KSK}.}}

\end{document}